\documentclass[paper,a4paper,11pt,twoside]{JHEP3}
\usepackage{psfrag}
\usepackage{epsfig}
\usepackage{amsmath}
\usepackage{multirow}
\usepackage{amssymb}
\usepackage{bbm}
\usepackage{axodraw}
\usepackage{cite}

\title{Hadronic form factors in Lattice QCD at
small and vanishing momentum transfer}

\author{%
J.M.~Flynn, A.~J\"uttner, C.T.~Sachrajda\\
School of Physics and Astronomy, University of Southampton,\\
Southampton, SO17 1BJ, UK.
}
\author{%
P.A.~Boyle, J.M.~Zanotti\\
School of Physics, The University of Edinburgh,\\
Edinburgh, EH9 3JZ, UK.}
\author{UKQCD Collaboration}

\preprint{%
Edinburgh 2007/6\\
SHEP-07-07\\
\today
}

\abstract{%
The introduction of partially twisted boundary conditions allows
weak and electromagnetic form factors to be evaluated at specified
values of the hadronic momenta (and hence momentum transfers) in
lattice simulations. We present and demonstrate this technique for
the computation of the $K\to \pi$ semileptonic form factor at zero
momentum transfer and for the electromagnetic form factor of the
pion at arbitrarily small momentum transfers. These exploratory
computations are carried out in full QCD with 3 flavours of sea
quarks, but with only two values of $m_u=m_d$ which limits our
ability to perform the chiral extrapolations. The results should
therefore be viewed primarily as a demonstration of the
feasibility of the method. For the $K\to \pi$ form factor we
compare the new technique to the conventional approach and for the
pion form factor we assess our results for very small momentum
transfer with the help of chiral perturbation theory.}


\begin{document}
%%%%%%%%%%%%%%%%%%%%%%%%%%%%%%%%%%%%%%%%%%%%%%%%%%%%%%%%%%%%%%%%%%%%%%%%%%%%%%%%
\section{Introduction}
In this paper we investigate the feasibility of using partially
twisted boundary conditions in lattice computations to evaluate
hadronic form factors at any chosen value of the momentum
transfer. With conventional periodic boundary conditions on the
quark fields on a spatial lattice of volume $L^3$ the components
of the hadrons' momenta are quantized to be integer multiples of
$2\pi/L$~\footnote{Throughout this paper, $L$ denotes the spatial
extent of the lattice.}. This leads to a very poor momentum
resolution for phenomenological studies. For example, on a $24^3$
lattice with lattice spacing $a\simeq 0.1$\,fm, the components of
the momentum are quantized in steps of about 0.5\,GeV\,. The
available momentum transfers are therefore also discrete. Using
partially twisted boundary conditions we show that it is indeed
possible to evaluate the form factors at any required momentum
transfer, provided of course that the hadronic momenta are
sufficiently small for lattice artefacts to be negligible. We
illustrate the technique by evaluating:
\begin{enumerate}
\item[i)] the $K\to\pi$ form factors at zero $q^2$, where $q$ is
the momentum transfer. These are required for the determination of
the $V_{us}$ matrix element of the CKM matrix. In the standard
approach to lattice computations of the $K\to\pi$ form factor at
$q^2=0$, first proposed by Becirevic et
al.~\cite{Becirevic:2004ya,Becirevic:2004bb} and subsequently
employed in a number of
simulations~\cite{Tsutsui:2005cj,Dawson:2006qc,Antonio:2006ev,
Antonio:2007mh}, the form factors are calculated (very precisely)
at $q^2=q^2_{\textrm{max}}$ (corresponding to the pion and kaon
both at rest) and somewhat less precisely at other accessible
values of $q^2$. The results are then interpolated to $q^2=0$. In
ref.\,\cite{Guadagnoli:2005be}, twisted boundary conditions were
used in a quenched study to obtain the form factors at values of
$q^2$ not accessible with periodic boundary conditions. In this
paper we demonstrate that by using partially twisted boundary
conditions in a dynamical simulation it is possible to evaluate
the form factors directly at $q^2=0$ with comparable errors to the
conventional approach but without the need for any interpolation
in the momentum transfer, thus removing one source of systematic
error~\cite{Antonio:2007mh}. \item[ii)] the electromagnetic form
factor of the pion at low momentum transfers (from which, for
example, the charge radius can be determined). In particular we
evaluate the form factor at values of $q^2$ below the minimum
value obtainable with periodic boundary conditions; this minimum
is given by $|q^2|=2m_\pi(m_\pi-\sqrt{m_\pi^2+(2\pi/L)^2})$. In
contrast to recent lattice studies
\cite{Bonnet:2004fr,Capitani:2005ce,Brommel:2006ww} this allows
therefore for a direct evaluation of the charge radius of the
pion.
\end{enumerate}
These are two phenomenologically important examples, but we stress
that the techniques can also be applied to a wide variety of
hadronic matrix elements. The primary aim of this paper is to
demonstrate the feasibility of the technique. We have therefore
used a restricted set of quark masses and hence have only a very
limited control of the chiral extrapolation. Having demonstrated
the effectiveness of the technique we will now undertake a
large-scale computation of the form factors.

The plan for the remainder of this paper is as follows. In the
next section we introduce all the necessary definitions,
correlation functions and ratios of correlation functions and then
detail the new approaches to compute the $K_{l3}$ scalar and pion
form factor. In particular, in section~\ref{subsec:sketch} we
explain why it is possible to use a general set of partially
twisted boundary conditions for the evaluation of the
electromagnetic form factor of the pion. Section
\,\ref{sec:numerical} contains the details of our numerical study
together with the results and finally in section~\ref{sec:concs}
we present a brief summary and outlook.

%%%%%%%%%%%%%%%%%%%%%%%%%%%%%%%%%%%%%%%%%%%%%%%%%%%%%%%%%%%%%%%%%%%%%%%%%%%%%%%%
\section{Description of the Technique}

The matrix element of the vector current between initial and final
states consisting of pseudoscalar mesons $P_i$ and $P_f$,
respectively, is in general decomposed into two invariant form
factors:
\begin{equation}\label{eq:gen_ff}
\langle P_f(p_f)|V_\mu| P_i(p_i)\rangle =
    f_{P_iP_f}^+(q^2)(p_i+p_f)_\mu +
    f_{P_iP_f}^-(q^2)(p_i-p_f)_\mu\,,
\end{equation}
where $q=p_i-p_f$ is the momentum transfer. For $K\to\pi$
semileptonic decays $V_\mu$ is the weak current $\bar{s}\gamma_\mu
u$, $P_i=K$ and $P_f=\pi$, whereas for the electromagnetic form
factor of the pion $V_\mu$ is the electromagnetic current, both
$P_i$ and $P_f$ are pions and vector current conservation implies
that $f_{\pi\pi}^-(q^2)=0$. The form factors $f_{P_iP_f}^+(q^2)$
and $f_{P_iP_f}^-(q^2)$ contain the non-perturbative QCD effects.
In addition to the matrix elements considered here, form factors
for other phenomenologically interesting semileptonic decays, for
example for $B\to\pi$, $B\to D$ and $D\to K$ decays as well as
those for decays into a vector final state, are also being
computed in lattice simulations. As explained in the introduction,
with the conventional periodic boundary conditions the form
factors can only be evaluated at values of $q^2$ such that the
components of momenta of the pseudoscalars, $\vec{p}_i$ and
$\vec{p}_f$, are integer multiples of $2\pi/L$. Our aim is to
compute the form factors at arbitrary preselected values of $q^2$
and in the following subsections we explain our technique for
achieving this. We start with a brief introduction to twisted
boundary conditions.

%%%%%%%%%%%%%%%%%%%%%%%%%%%%%%%%%%%%%%%%%%%%%%%%%%%%%%%%%%%%%%%%%%%%%%%%%%%%%%%%
\subsection{Twisted Boundary Conditions}

It is well known that the choice of boundary conditions for
particles or fields in quantum mechanics or field theory in a
finite volume governs the momentum spectrum. This observation has
been exploited in many applications. More recently it has been
appreciated that by varying the boundary conditions the momentum
resolution for lattice phenomenology can be significantly
improved~\cite{Bedaque:2004kc,deDivitiis:2004kq,Sachrajda:2004mi,
Bedaque:2004ax,Tiburzi:2005hg,Flynn:2005in,Guadagnoli:2005be,
Aarts:2006wt,Tiburzi:2006px,Bunton:2006va}. However, if a new
simulation (i.e. a new set of gauge configurations) were necessary
for every choice of momentum, the use of twisted boundary
conditions would be prohibitively expensive in computing resources
and therefore impracticable. In ref.~\cite{Sachrajda:2004mi} it
was demonstrated that for processes without final state
interactions, such as the form factors studied in this paper, it
is sufficient to apply twisted boundary conditions only on the
valence quarks, whilst using sea quarks defined with periodic
boundary conditions (see also ref.\cite{Bedaque:2004ax}). In this
way the need for new simulations is avoided and the method becomes
practicable. The introduction of such \textit{partially twisted}
boundary conditions changes the finite-volume corrections, but, as
demonstrated in refs.\,\cite{Sachrajda:2004mi,Jiang:2006gn}, they
remain exponentially small in the volume and, as is standard, we
neglect them.

In our study we use partially twisted boundary conditions,
combining gauge field configurations generated with sea quarks
with periodic boundary conditions with valence quarks with twisted
boundary conditions, i.e. the valence quarks satisfy
\begin{equation}
\psi(x_k+L)=e^{i\theta_k}\psi(x_k),\qquad(k=1,2,3)\,,
\end{equation}
where $\psi$ is either a strange quark $s$ or a light quark $q$.
By varying $\vec{\theta}$ we can tune the momenta of the mesons
continuously. For the purposes of our study it will be
sufficient to twist only
the valence quark in each meson, with the valence antiquark
satisfying periodic boundary conditions (the generalization to
antiquarks with twisted boundary conditions is also
straightforward). The dispersion relation for the mesons is then
\begin{equation}
 E = \sqrt{m^2 + (\vec p_{\rm FT} + \vec \theta/L)^2},
\end{equation}
where $m$ is the mass of the meson and $\vec p_{\rm FT}$ is the
meson momentum induced by Fourier summation (the components of
$\vec p_{\rm FT}$ are integer multiples of $2\pi/L$).

For the matrix element in (\ref{eq:gen_ff}) with the initial and
the final meson carrying momenta $\vec p_i$ and $\vec p_f$,
respectively, the momentum transfer between the initial and the
final state meson is
\begin{equation}\label{eq:mom_transfer}
q^2=(p_i-p_f)^2=\left\{[E_i(\vec p_i)-E_f(\vec p_f)]^2
        -\left[(\vec p_{{\rm FT},i}+\vec{\theta}_i/L)
          -(\vec p_{{\rm FT},f}+\vec{\theta}_f/L)\right]^2\right\}.
\end{equation}

%%%%%%%%%%%%%%%%%%%%%%%%%%%%%%%%%%%%%%%%%%%%%%%%%%%%%%%%%%%%%%%%%%%%%%%%%%%%%%%%
\subsection{Correlation Functions}

In order to determine the form factors we compute two- and
three-point correlation functions. The two-point function is
defined by
\begin{equation}
C_i(t,\vec p_i)=\sum_{\vec{x}}e^{i\vec{p}_i\cdot\vec{x}} \langle
\,O_i(t,\vec{x})\, O_i^\dagger(0,\vec{0})\,\rangle
    =\frac{
            |Z_i|^2}{2E_i}
            \left(e^{-E_it}+ e^{-E_i(T-t)}\right)\, ,
\label{eq:twopt}\end{equation} where $i=\pi$ or $K$, the $O_i$ are
local pseudoscalar interpolating operators for the corresponding
mesons $O_\pi= \bar q \gamma_5 q$ and $O_K=\bar s\gamma_5 q$ and
we assume that $t$ and $T-t$ (where $T$ is the temporal extent of
the lattice) are sufficiently large that the correlation function
is dominated by the lightest state (i.e. the pion or kaon). The
constants $Z_i$ are given by
$Z_i=\langle\,P_i\,|\,O_i^\dagger(0,\vec{0})\,|\,0\,\rangle$\,. The
three-point functions are defined by
\begin{eqnarray}
C_{P_iP_f}(t_{i},t,t_{f},\vec p_i,\vec p_f) &=&
\sum_{\vec{x}_f,\vec{x}} e^{i\vec{p}_f\cdot(\vec{x}_f-\vec{x})}
e^{i\vec{p}_i\cdot\vec{x}} \langle\, O_f(t_{f},\vec x_f)\,
V_4(t,\vec{x})\,O_i^\dagger(t_{i},\vec 0)\,\rangle\nonumber\\%[2mm]
    &=&
        \frac{Z_i\,Z_f}{4E_iE_f}\, \langle\,P_f(\vec{p}_f)\,|\,V_4(0)\,|\,
        P_i(\vec{p}_i)\,\rangle\,\nonumber\\ &&\hspace{-0.8in}
        \times\left\{\theta(t_f-t)\,e^{-E_i(t-t_i)-E_f(t_f-t)}\ -\
\theta(t-t_f)\,e^{-E_i(T+t_i-t)-E_f(t-t_f)}\right\}\,,\label{eq:threept}
\end{eqnarray}
where $P_{i,f}$ is a pion or a kaon and $V_4$ is the time
component of the vector current with flavour quantum numbers to
allow the $P_i\to P_f$ transition and where we have defined
$Z_f=\langle\, 0\,|\,O_f(0,\vec 0)|\,P_f\,\rangle$. Again we
assume that all the time intervals are sufficiently large for the
lightest hadrons to give the dominant contribution.
For the remainder of this paper we choose to keep $t_{i}$ and $t_{f}$
fixed (with $t_i=0$ and $t_f=T/2$) and we will therefore only
explicitly refer to them where necessary\,\footnote{ In practice we
average over the results obtained with different origins of the
quark propagators. In these cases all the coordinates given here
have to be translated accordingly.}
and write $C_{P_iP_f}(t,\vec{p}_i,\vec{p}_f)$ with
just three arguments.

The correctly normalized vector currents $V_\mu$ in
eq.\,(\ref{eq:threept}) are obtained by multiplying the local
currents used in the numerical simulations by the normalization
constant $Z_V$ defined by
\begin{equation}\label{eq:zv}
Z_V = \frac{1}{2}\frac{C_{\pi} (t_f=T/2,0)}
    {C^B_{\pi\pi}(t,0,0) },
\end{equation}
where the index $B$ in $C^B_{\pi\pi}(t,0,0)$ implies that the bare
vector current is being used. The factor of $\frac12$ in
eq.\,(\ref{eq:zv}) corresponds to the two terms on the right hand
side of eq.\,(\ref{eq:twopt}).

In order to extract the matrix element
$\langle\,P_f(\vec{p}_f)\,|\,V_4(0)\,|\,P_i(\vec{p}_i)\,\rangle$
effectively, it is convenient to define the three ratios:
\begin{equation}\label{eq:ratios}
\begin{array}{rcl}
R_{1,\,P_iP_f}(\vec{p}_i,\vec{p}_f)&{=}&
 4\sqrt{E_i E_f}\, \sqrt{\frac
 {C_{P_iP_f}(t,\vec p_i,\vec p_f)\,C_{P_fP_i}(t,\vec p_f,\vec p_i)}
 {C_{P_i}(T/2,\vec p_i)\,C_{P_f}(T/2,\vec p_f)}},
 \\[4mm]
R_{2,\,P_iP_f}(\vec{p}_i,\vec{p}_f)&=&
 2\sqrt{{E_i E_f }}\,
 \sqrt{
 \frac{C_{P_iP_f}(t,\vec p_i,\vec p_{f})
        \,C_{P_fP_i}(t,\vec p_f,\vec p_{i})}
 {C_{P_iP_i}(t,\vec p_i,\vec p_{i})\,
    C_{P_fP_f}(t,\vec p_f,\vec p_{f})}},\\[4mm]
R_{3,\,P_iP_f}(\vec{p}_i,\vec{p}_f)&=&
 4{\sqrt{E_i E_f}}\,
\frac{C_{P_iP_f}(t,\vec p_i,\vec p_{f})}{C_{P_f}(T/2,\vec p_f)}\,
    \sqrt{
    \frac{C_{P_i}(T/2-t,\vec p_i)\,C_{P_f}(t,\vec p_f)\,C_{P_f}(T/2,\vec p_f)}
    {C_{P_f}(T/2-t,\vec p_f)\,C_{P_i}(t,\vec p_i)\,C_{P_i}(T/2,\vec
    p_i)}}\,.
\end{array}
\end{equation}
For sufficiently large $t$ and $T/2-t$, so that only the lightest
mesons contribute significantly to each of the correlation
functions, each of the three ratios is independent of $t$ and is
equal to the matrix element
$\langle\,P_f(\vec{p}_f)\,|\,V_4(0)\,|\,P_i(\vec{p}_i)\,\rangle$.
Here we are assuming that $t$ is in the forward half of the
lattice $0<t<T/2$. The correlation functions for $t$ in the
backward half, $T/2<t<T$ are readily related to those in the
forward half and hence can be combined with them to construct the
ratios in (\ref{eq:ratios}). We discuss the quality of the
plateaus and the numerical determination of the form factors in
section~\ref{sec:numerical}.

%***********************************

\subsection{The Pion's Form Factor with Twisted Boundary
Conditions}\label{subsec:sketch}

A sketch of the quark-flow diagram for the transition in
eq.\,(\ref{eq:gen_ff}), with the final-state meson $P_f$ composed
of valence quarks ($q_1\bar{q}_3$) and the initial-state meson
with valence quarks ($q_2\bar{q}_3$) is as follows:
\begin{center}\begin{picture}(120,60)(-60,-30) \ArrowLine(-50,0)(-25,0)
\ArrowLine(25,0)(50,0)\Oval(0,0)(12,25)(0)
\GCirc(-25,0){3}{0.5}\GCirc(25,0){3}{0.5} \GCirc(0,12){3}{0.5}
\Text(-19,12)[b]{$q_2$}\Text(19,12)[b]{$q_1$}
\Text(0,-15)[t]{$q_3$}\Text(0,17)[b]{$V_\mu$}\Text(-54,0)[r]{$P_i$}
\Text(54,0)[l]{$P_f$}\ArrowLine(0.5,-12)(-0.5,-12)
\end{picture}\end{center}
For $K_{\ell 3}$ decays, specifically for the decay
$\bar{K}^0\to\pi^+\ell\nu_l$, each of the three valence has a
different flavour, $q_1=u,\,q_2=s$ and $q_3=d$, and the partially
twisted theory can be readily constructed as discussed in
ref.\,\cite{Sachrajda:2004mi}. We can therefore introduce three
independent twisting angles for the three flavours. For the
electromagnetic form factor of the pion however, $q_1$ has the
same flavour as $q_2$, nevertheless it is still possible to use
partially twisted boundary conditions to evaluate the form factor,
with three different twisting angles for the three valence quarks,
as we now explain~\footnote{In the numerical work described in
section\,\ref{sec:numerical} we choose to keep the twisting angle
of the spectator quark ($\vec\theta_3$) equal to zero.}.
\begin{enumerate}
\item[(a)] We start by imagining that we evaluate the matrix
element $\langle\pi(p_f)\,|\,V_\mu\,|\,\pi(p_i)\rangle$ in an
infinite volume in full QCD with 3 flavours of sea quarks. We
assume isospin symmetry, and it will be important to note that in
this case $G$-parity implies that only the isovector component of
the electromagnetic current couples to pions. When we consider
partially quenched QCD below, this will imply that the vector
current is composed of valence quark fields. \item[(b)] We next
consider the partially quenched 3-flavour theory in which
$m_u^V=m_d^V=m_u^S=m_d^S=m_s^V$, where the superscripts $V$ and
$S$ label \textit{valence} and \textit{sea}, respectively, and
$m_s^S$ is equal to the physical strange quark mass. The partial
quenching arises because the mass of the valence strange quark is
different to that of the sea strange quark. However, since the
valence strange quark plays no role in the evaluation of
$\langle\pi(p_f)\,|\,V_\mu\,|\,\pi(p_i)\rangle$, this matrix
element is correctly given in this theory. \item[(c)] We now
exploit the $SU(3)$ flavour symmetry in the valence sector which
implies that, for example,
\begin{equation}
\langle\pi^+(p_f)\,|\,\bar{u}\gamma^\mu u\,|\,\pi^+(p_i)\rangle=
-\langle\pi^+(p_f)\,|\,\bar{d}\gamma^\mu d\,|\,\pi^+(p_i)\rangle=
\langle\pi^+(p_f)\,|\,\bar{u}\gamma^\mu
s\,|\,\bar{K}^0(p_i)\rangle\,,
\end{equation}
where the fields in the currents correspond to valence quarks. The
pion's form factor in this partially quenched theory is therefore
equal to that of the $K\to\pi$ transition (the degeneracy of the
three flavours of valence quark implies that there is now also
only a single form factor also for $K\to\pi$ transitions).

Up to this point we have not made any approximations. For example,
the $SU(3)$ flavour symmetry in the valence sector implies that
exactly the same diagrams arise in chiral perturbation theory
($\chi$PT) in the evaluation of the matrix element\linebreak
$\langle\pi^+(p_f)\,|\,\bar{u}\gamma^\mu
s\,|\,\bar{K}^0(p_i)\rangle$ in the partially quenched theory as
for $\langle\pi(p_f)\,|\,V_\mu\,|\,\pi(p_i)\rangle$. Of course the
labelling of the quark content of the mesons may be different; in
one case we may have a meson with a strange valence quark and in
the other an $u$-quark, but as they are degenerate the diagrams
give identical contributions (as they must by symmetry). In
particular, in contrast to the general case for partially quenched
theories, there are no \textit{hairpin} contributions here.
\item[(d)] Finally we consider performing the simulations in
finite volume, taking all the sea quarks to have periodic boundary
conditions. For the three flavours of valence quarks however, we
introduce different twists, $\vec{\theta}_u,\,\vec{\theta}_d$ and
$\vec{\theta}_s$, which changes the momentum spectrum but not the
mass spectrum. For propagators in $\chi$PT with quark content of
the form $\bar{q}q\ (q=u^V,d^V,s^V)$, the effects of the twisted
boundary condition cancel and the spectrum is the same as for the
mesons composed of the corresponding sea quarks (i.e. as if the
boundary conditions were periodic). Thus, in spite of the
different masses of the valence and sea strange quarks, no
double-pole hairpin contributions are introduced.

For valence flavour non-singlet mesons, the momentum spectrum is
changed by using twisted boundary conditions rather than periodic
ones. However, as explicitly demonstrated in Appendix A of
ref.\,\cite{Sachrajda:2004mi}, at one-loop order in $\chi$PT the
resulting summations in finite-volume are equal to the
corresponding infinite volume integrals, up to exponentially small
terms in the volume. Such exponentially small terms are also
present with periodic boundary conditions and are generally
neglected. \end{enumerate} We have therefore established that we
can evaluate the pion form factor with 3 different twisting angles
for the valence quarks (up to the usual exponential precision in
the volume).

%%%%%%%%%%%%%%%%%%%%%%%%%%%%%%%%%%%%%%%%%%%%%%%%%%%%%%%%%%%%%%%%%%%%%%%%%%%%%%%%
\subsection{\boldmath{$f^{+}_{K\pi}(0)$} and \boldmath{$f_{\pi\pi}(q^2)$} at Small
Momentum Transfers}

We wish to compute the scalar form factor for $K_{\ell 3}$ decays
at zero momentum transfer, $f^0_{K\pi}(0)$. The scalar form factor
is defined in terms of $f^+_{K\pi}$ and $f^-_{K\pi}$ by:
\begin{equation}\label{eq:define_f0}
f_{K\pi}^0(q^2)=f_{K\pi}^+(q^2)+{q^2\over m_K^2
-m_\pi^2}f_{K\pi}^-(q^2),
\end{equation}
and $f^0_{K\pi}(0)=f_{K\pi}^+(0)$.

Our approach is to use twisted boundary conditions to induce
momenta for the pion and kaon such that $q^2=0$. A simple way to
do this is to take the pion (kaon) to be at rest and to tune the
momentum of the kaon (pion). We therefore compute the ratios
\begin{equation}\label{eq:twists}
 \begin{array}{llcccc}
&  R_{\alpha,K\pi}(\vec{p}_K,\vec{0})&\textrm{with}&
  |\vec{\theta}_K| =
           L\sqrt{({m_K^2+m_\pi^2 \over 2m_\pi})^2 - m_K^2}
      &\textrm{and}&\vec{\theta}_\pi=\vec{0}\\[2mm]
{\rm and}&  R_{\alpha,K\pi}(\vec{0},\vec{p}_\pi)&\textrm{with}&
          |\vec{\theta}_\pi| =L
          \sqrt{({m_K^2+m_\pi^2 \over 2m_K})^2 -
          m_\pi^2}&\textrm{and}
      &\vec{\theta}_K =\vec{0}\,,\\[2mm]
 \end{array}
\end{equation}
where $\alpha=1,2,3$. The momenta of the mesons are given by
$\vec{p}_K=\vec{\theta}_K/L$ and $\vec{p}_\pi=\vec{\theta}_\pi/L$
and it can be readily verified that the choices of twisting angles
in the two lines of eq.(\ref{eq:twists}) both correspond to
$q^2=0$.

The required form factor, $f^0_{K\pi}(0)$, can be obtained
directly from a linear combination of the ratios in
eq.\,(\ref{eq:twists}):
\begin{equation}\label{eq:lin_comb}
f_{K\pi}^0(0)=\frac{ R_{\alpha,K\pi}(\vec{p}_K,\vec{0})(m_K-E_\pi)
- R_{\alpha,K\pi}(\vec{0},\vec{p}_\pi)(E_K-m_\pi) }{
(E_K+m_\pi)(m_K-E_\pi)-(m_K+E_\pi)(E_K-m_\pi)
}\qquad(\alpha=1,2,3)\,.
\end{equation}
Here $E_K$ ($E_\pi$) is the energy of the kaon (pion)
corresponding to the momentum induced by the twisting angle in the
first (second) line of eq.(\ref{eq:twists}).

By using the spatial component of the vector current $V_k$
($k=1,2$ or 3) it is possible to tune the momenta such that
$q^2=0$ and $q_k=0$ so that one obtains the form-factor $f^+(0)$
directly. We find however, that this procedure leads to a
significantly larger statistical error.

The case of the pion's electromagnetic form factor is simpler
since current conservation implies that $f_{\pi\pi}^-(q^2)=0$ so
that (dropping the redundant superscript $+$)
\begin{equation}
 \langle \pi^+(p_f)|V_\mu(0)|\pi^+(p_i)\rangle =f_{\pi\pi}(q^2)\,(p_i+p_f)_\mu.
\end{equation}
$f_{\pi\pi}(q^2)$ can therefore be directly computed from the
ratios in eq.(\ref{eq:ratios}) by inducing the required momenta
for the initial- and final-state pions.

%%%%%%%%%%%%%%%%%%%%%%%%%%%%%%%%%%%%%%%%%%%%%%%%%%%%%%%%%%%%%%%%%%%%%%%%%%%%%%%%
\subsection{Comparison with the Conventional
Approaches}\label{subsec:conventional}

One aim of this paper is to compare the precision with which we
can determine the form factors using the techniques introduced in
the preceding subsection with that obtained using standard
methods. The numerical comparison will be given in
section~\ref{sec:numerical}, here we describe what the
conventional approaches are.

Recent calculations of  $f^0_{K\pi}(0)$ follow the procedure
introduced by Becirevic et al.~\cite{Becirevic:2004ya}. The scalar
form factor at $q^2=0$ is determined using a phenomenologically
motivated interpolation  of $f^0_{K\pi}(q^2)$ between the point at
$q^2_{\rm max} = (m_K-m_\pi)^2$ and points at negative values of
$q^2$ which are accessible by Fourier summation with periodic
boundary conditions. The value of $f^0_{K\pi}(q^2_{\rm max})$ is
readily obtained with excellent statistical precision from
\begin{equation}\label{Eqn:R_define}
R_{2;K\pi}(\vec{0},\vec{0}) = f_{K\pi}^0(q^2_{\rm max})(m_K+m_\pi)\,.
\end{equation}
The interpolation in $q^2$ (to $q^2=0$) is constrained by
computing $f_{K\pi}^{+}(q^2)$ and $f_{K\pi}^{-}(q^2)$ (and hence
$f_{K\pi}^{0}(q^2)$) for a variety of values of $q^2$. In order to
improve precision, where possible ratios of correlation functions
are used. $f_{K\pi}^{+}(q^2)$ and $f_{K\pi}^{-}(q^2)$ can be
written in the form
\begin{equation}\label{eq:conv_mainformula}
f_{K\pi}^+(q^2)={F(p_K,p_\pi)}f_{K\pi}^0(q^2_{\rm max}) \left(
1+{E_K({\vec p}_K)-E_\pi({\vec p}_\pi)\over E_K({\vec p}_K)+
        E_\pi({\vec p}_\pi)}
{\xi(q^2)}
\right)^{-1},
\end{equation}
and
\begin{equation}
f_{K\pi}^-(q^2)=f_{K\pi}^+(q^2) {\xi(q^2)}.
\end{equation}
In order to define the quantities $F(p,p^\prime)$ and $\xi(q^2)$
used in these expressions we start by defining the ratio of
correlation functions
\begin{equation}
\tilde R_k(\vec{p}_K,\vec{p}_\pi)=\frac {C_{k,\,K\pi}(t,{\vec
p}_K,{\vec p}_\pi)\,
 C_{KK}(t,{\vec p}_K,{\vec p}_\pi)}
{C_{K\pi}(t,{\vec p}_K,{\vec p}_\pi)\,
 C_{k,\,KK}(t,{\vec p}_K,{\vec p}_\pi)}\,,
\end{equation}
where $k=1,2,3$ is a spatial index and $C_{k,\,P_iP_f}(t,{\vec
p}_i,{\vec p}_f)$ is the three point correlation function defined
in eq.\,(\ref{eq:threept}) with $V_4$ replaced by
$V_k$~\footnote{Since this is the only place where we require the
spatial component of the vector current, for simplicity of
notation we do not introduce the Lorentz index in
eq.~(\ref{eq:threept}). Note however that the minus sign in
eq.~(\ref{eq:threept}) becomes a plus sign for spatial indices of
the vector current. $C_{P_iP_f}$ implicitly corresponds to the
matrix element of $V_4$.}. For time intervals such that only the
lightest states contribute significantly, $\tilde{R}_k$ is
independent of $t$. $\xi$ and $F$ are then defined by:
\begin{equation}\label{Eqn:xi_define}
\xi(q^2)=\frac { -\left(E_K({\vec p}_K)+E_K  ({\vec
p}_\pi)\right)\left(p_K+p_\pi\right)_k +\left(E_K({\vec
p}_K)+E_\pi({\vec p}_\pi)\right)\left(p_K+p_\pi\right)_k
    \tilde R_k(\vec{p}_K,\vec{p}_\pi)
} { \hspace{4mm}
 \left(E_K({\vec p}_K)+E_K  ({\vec p}_\pi)\right)\left(p_K-p_\pi\right)_k
-\left(E_K({\vec p}_K)-E_\pi({\vec
p}_\pi)\right)\left(p_K+p_\pi\right)_k
        \tilde R_k(\vec{p}_K,\vec{p}_\pi)
},
\end{equation}
and
\begin{equation}\label{Eqn:F_define}
\frac {C_{K\pi}(t,{\vec p}_K,{\vec p}_\pi)
 \,C_{K}(t,{\vec 0})
 \,C_{\pi}(T/2-t,{\vec 0})}
{C_{K\pi}(t,{\vec 0},{\vec 0})
 \,C_{K}(t,{\vec p}_K)
\, C_{\pi}(T/2-t,{\vec p}_\pi)}= \frac{E_K({\vec p}_K)+E_\pi({\vec
p}_\pi)}{m_K+m_\pi}F(p_K,p_\pi)\,.
\end{equation}
Having determined $f_{K\pi}^+(q^2)$ and $f_{K\pi}^-(q^2)$ for a
variety of values of $q^2<0$ and at $q^2_{\textrm{max}}$, we fit
the results to some ansatz for the $q^2$ behaviour and determine
the scalar form factor at $q^2=0$. It should be noted that in
addition to the systematic uncertainties introduced by the choice
of ansatz, one is limited to the number of values of $q^2$ one can
use while keeping the lattice artefacts small enough for the
required precision.

Recent examples for the use of this approach have been presented
in
refs.\,\cite{Tsutsui:2005cj,Dawson:2006qc,Antonio:2006ev,Antonio:2007mh}.
%%%%%%%%%%%%%%%%%%%%%%%%%%%%%%%%%%%%%%%%%%%%%%%%%%%%%%%%%%%%%%%%%%%%%%%%%%%%%%%%
\section{Results for the Form Factors}\label{sec:numerical}

In this section we describe the details of our numerical
simulation and present our results.

%%%%%%%%%%%%%%%%%%%%%%%%%%%%%%%%%%%%%%%%%%%%%%%%%%%%%%%%%%%%%%%%%%%%%%%%%%%%%%%%
\subsection{Lattice Parameters}
For the numerical studies presented
in this paper we  use two ensembles
out of the set of $N_f=2+1$ flavour Domain Wall
Fermion~\cite{Kaplan:1992bt,Shamir:1993zy,Furman:1994ky}
configurations with $(L/a)^3\times T/a\times L_s=16^3\times
32\times 8$ which were jointly generated by the UKQCD/RBC
collaborations using the QCDOC computer
\cite{qcdoc1,qcdoc2,qcdoc3,qcdoc4}. A detailed study of the
light-hadron spectrum and other hadronic quantities using these
configurations has recently been reported in
ref.\,\cite{Antonio:2006px}. In particular, we use the gauge
configurations generated with the DBW2  gauge
action~\cite{Takaishi:1996xj,deForcrand:1999bi} at $\beta=0.72$.
The bare strange quark mass is $am_s=0.04$ and we use two
different ensembles with light quark masses $am_l=0.02$ and
$am_l=0.01$ respectively. The corresponding pion and kaon masses
are summarized in table \ref{tab:kin} and for the inverse lattice
spacing we take $a^{-1}=1.6(1)$\,GeV. We use the jackknife
technique to estimate the statistical errors.
%%%%%%%%%%%%%%%%%%%%%%%%%%%%%%%%%%%%%%%%%%%%%%%%%%%%%%%%%%%%%%%%%%%%%%%%%%%%%%%%

We generate the three point functions of type (\ref{eq:threept})
by contracting propagators \linebreak
$S(t,\vec x;t_i,\vec 0)$ from the origin to any point $(t,\vec x)$
 with the generalized quark
propagator~\cite{Martinelli:1988rr} defined by
\begin{equation}
S^\prime(t_i,\vec 0;t_f,\vec {p}_f;t,\vec{x})
    =\sum_{\vec{x}_f}\gamma_5\left(S(t,\vec {x};t_f,\vec{x}_f)
        \gamma^5 S(t_f,\vec{x}_f;t_i,\vec 0)\,
        e^{-i\vec{p}_f\cdot\vec{x}_f}\right)^\dagger \gamma_5\,,
\label{eq:sprimedef}\end{equation} where we suppress the label
indicating the twisting angle\footnote{ Note that the dagger in
the extended propagator reverses the sign of the twisting angle.}
on the propagators (cf. fig.\,\ref{fig:threepoint}). \FIGURE[t]{
\begin{minipage}{\linewidth}
\begin{center}
\psfrag{77}[c][t][1][0]{\small
${S^\prime}(t_i,\vec 0;t_f,\vec {p}_f;t,\vec{x})$}
\psfrag{88}[l][t][1][0]{\small$S(t,\vec x;t_i,\vec 0)$}
\psfrag{1}[t][b][1][0]{$q\,; \vec \theta=0$}
\psfrag{2}[l][t][1][0]{$q\, {\rm or}\, s;\vec \theta_1$}
\psfrag{3}[r][t][1][0]{$q\, {\rm or}\, s;\vec \theta_2$}
\psfrag{4}[c][t][1][0]{$V_\mu(t,\vec x)$}
\psfrag{5}[c][t][1][0]{$O_i(t_i,\vec 0)$}
\psfrag{6}[c][t][1][0]{$O_f(t_f,\vec x_f)$}
\epsfig{scale=.6,file=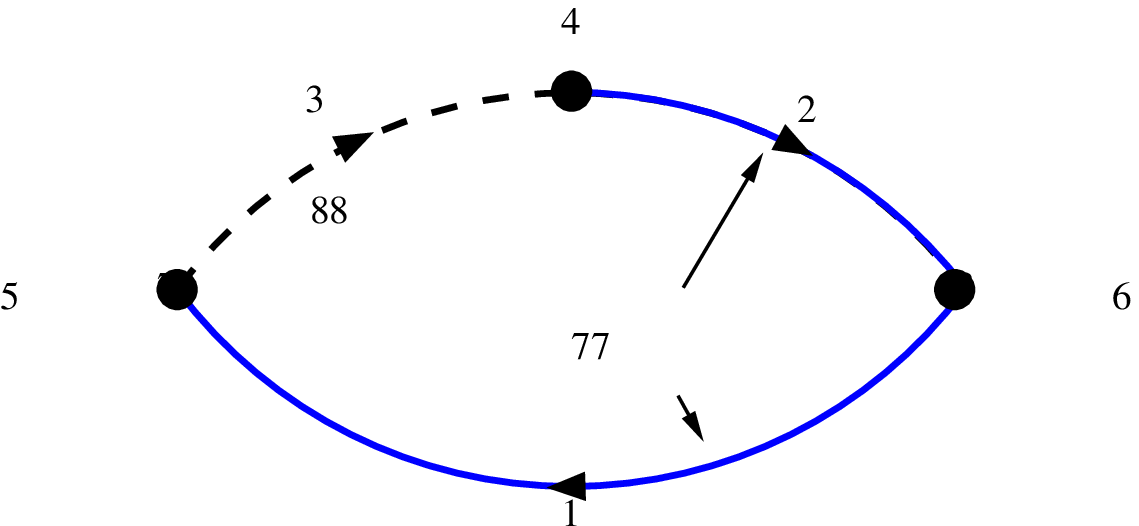}
\end{center}
\end{minipage}\\[3mm]
\caption{The three point function $C_{P_iP_f}(t_{i},t,t_{f},\vec p_i,\vec p_f)$
defined in (\ref{eq:threept}) in terms of the the quark propagator
$S(t,\vec x;t_i,\vec 0)$ (dashed black) and the generalized quark propagator
$S^\prime(t_i,\vec 0;t_f,\vec p;t,\vec x)$ (solid blue). }\label{fig:threepoint}
}
For the present study we generate the generalized propagators in
(\ref{eq:sprimedef}) with $\vec{p}_f=0$, although it would be
straightforward to extend the study to include other values.

For the $K_{\ell 3}$ form factors in all cases one meson is at
rest and the other has a momentum which is induced entirely by the
twisted boundary conditions. As a result we can evaluate all three
ratios $R_{\alpha,K\pi}(\vec{p}_K,\vec{p}_\pi)$ at a comparable
computational cost. A selection of the plateaus for the $K_{l3}$
decay is presented for illustration in fig.\,\ref{fig:comp_rat}.
We find that, for the choice of parameters used in this study,
ratio $R_{1,K\pi}(\vec p_K,\vec p_\pi)$ and $R_{2,K\pi}(\vec
p_K,\vec p_\pi)$ have the most pronounced plateaus and the fits
lead to comparable final statistical errors. The results we
present in the following have been obtained from fits to
$R_{1,K\pi}(\vec p_K,\vec p_\pi)$.

For the electromagnetic form factor of the pion we allow one of
the pions to have a non-zero value of Fourier momentum (i.e. its
momentum is given by $(2\pi\,\vec{n}+\vec\theta)/L$, where
$\vec{n}$ is a vector of integers and $\vec\theta$ is the vector
of twisting angles). In order to evaluate the ratios
$R_{1,\pi_i\pi_f}(\vec p_{\pi_i},\vec p_{\pi_f})$ and
$R_{2,\pi_i\pi_f}(\vec p_{\pi_i},\vec p_{\pi_f})$ we would require
the generalized propagators in (\ref{eq:sprimedef}) at non-zero
$\vec{p}$ and here we restrict our computations to the evaluation
of $R_{3,\pi_i\pi_f}(\vec p_{\pi_i},\vec p_{\pi_f})$. A typical
result for this ratio is illustrated in
fig.\,\ref{fig:comp_rat_pion}.

\FIGURE{ \psfrag{title}[c][t][1][0]{\scriptsize $R_1$}
\psfrag{xlabel}[t][t][1][0]{\scriptsize$t/a$}
\epsfig{scale=.19,angle=-90,file=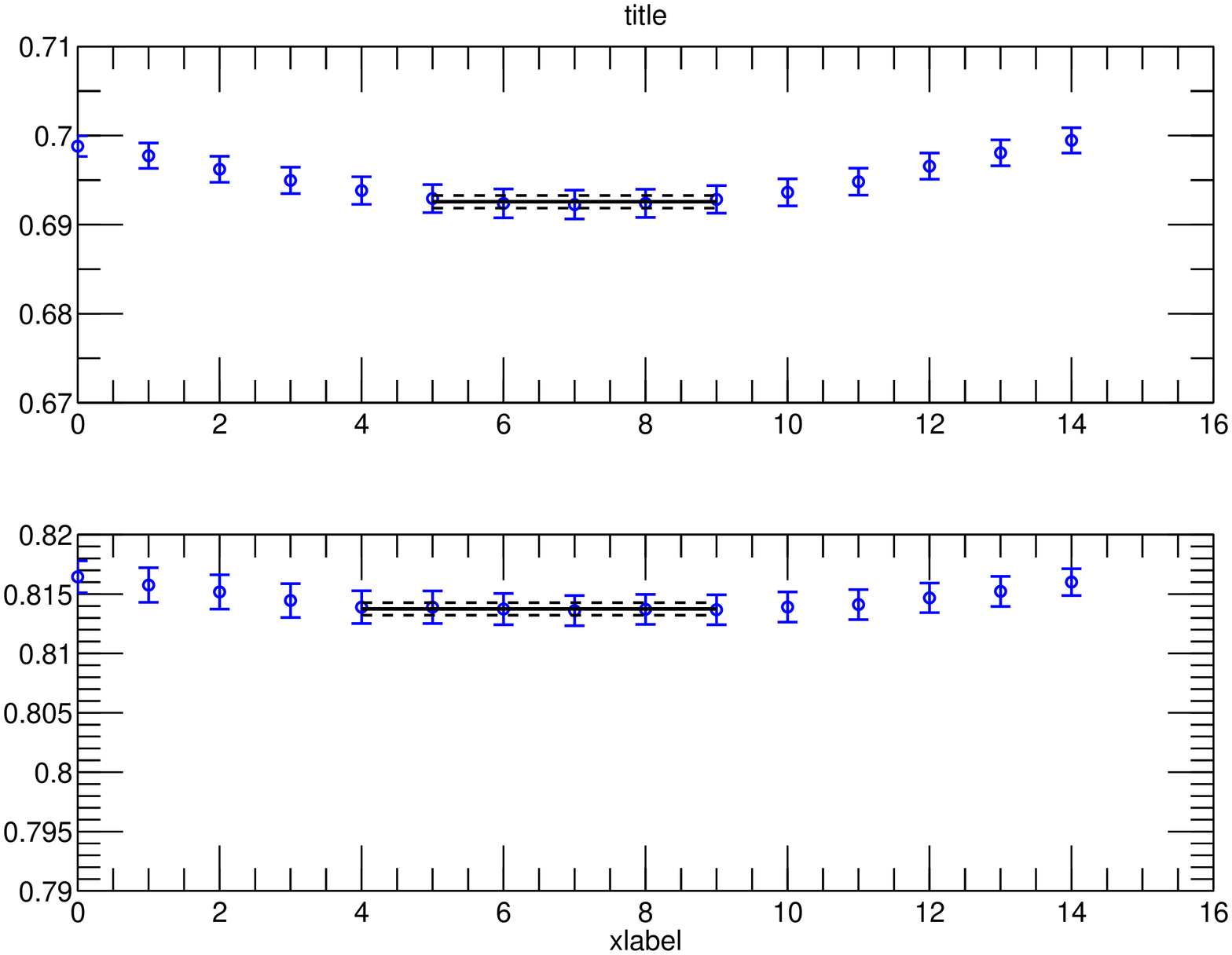}
\psfrag{title}[c][t][1][0]{\scriptsize$R_2$}
\epsfig{scale=.19,angle=-90,file=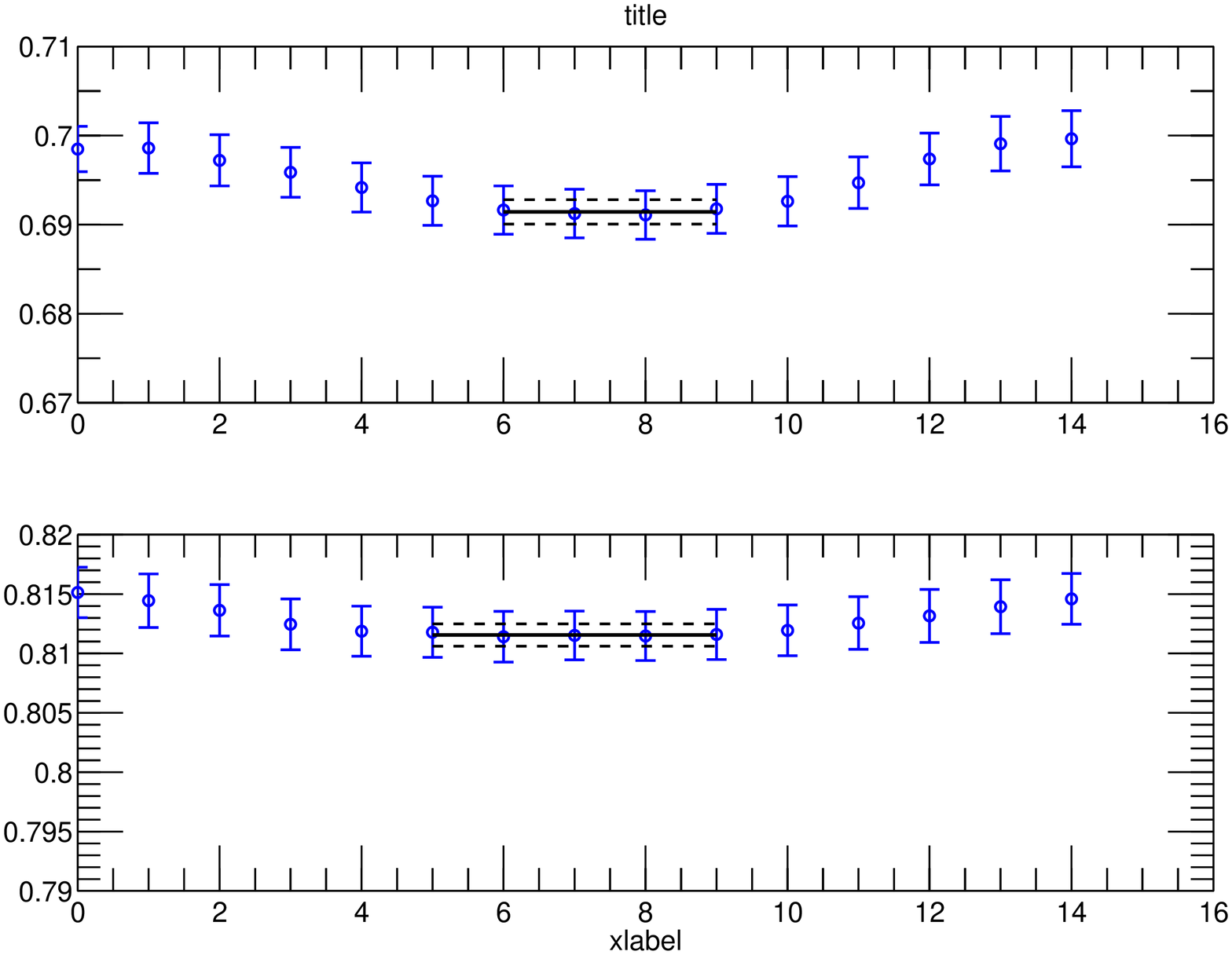}
\psfrag{title}[c][t][1][0]{\scriptsize$R_3$}
\epsfig{scale=.19,angle=-90,file=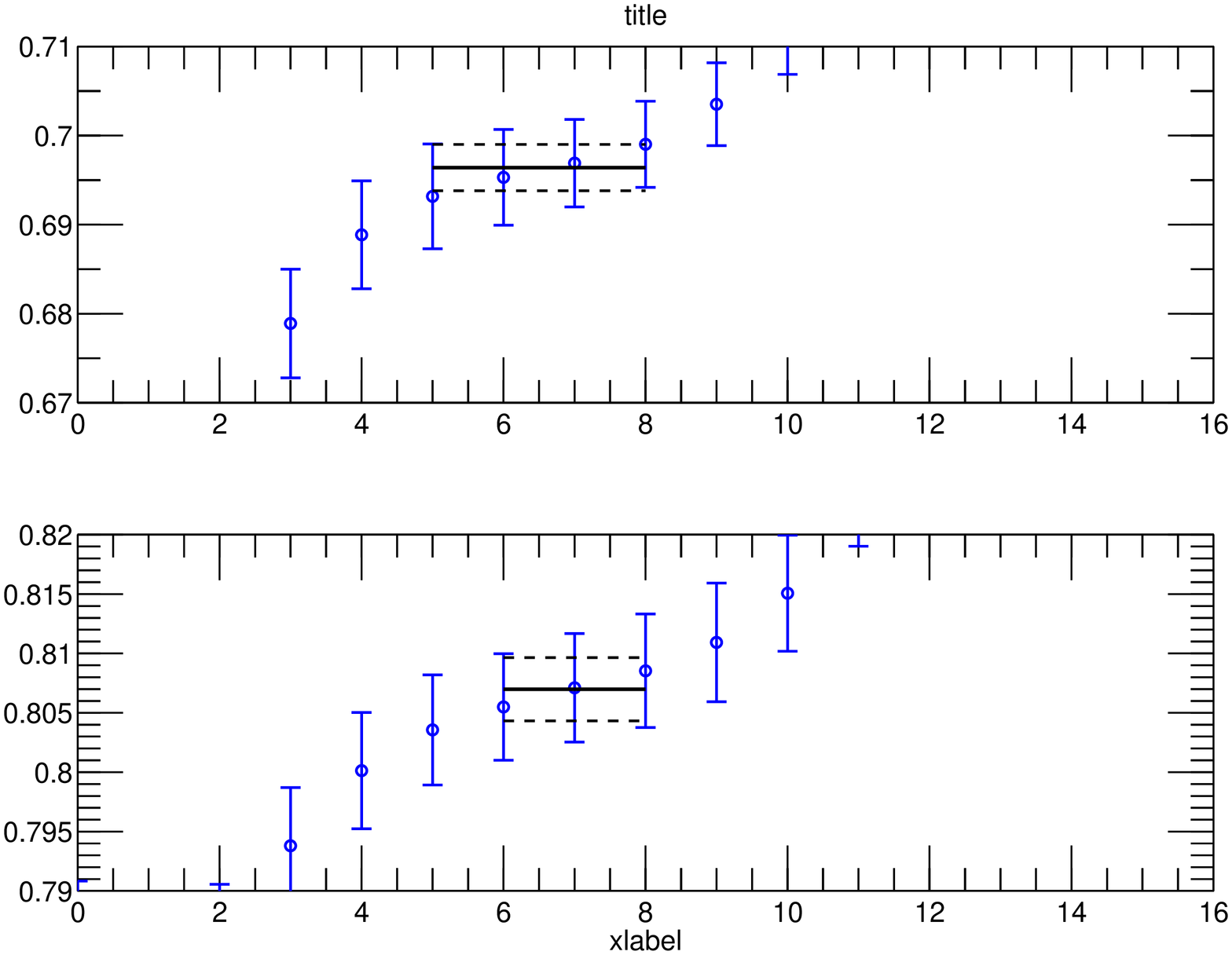}
\caption{Examples of results for the three different ratios
defined in (\ref{eq:ratios}) for the $K_{l3}$ decay. The two rows
correspond to the two quark masses $am_l=0.01$ (upper row) and
$am_l=0.02$ (lower row),
 respectively. In each case the kaon is at rest and
the pion has a momentum induced by the twisting angle. }
\label{fig:comp_rat} } \FIGURE{
\psfrag{title}[c][t][1][0]{\scriptsize$R_3$}
\psfrag{t/a}[t][t][1][0]{\scriptsize$t/a$}
\epsfig{scale=.19,angle=-90,file=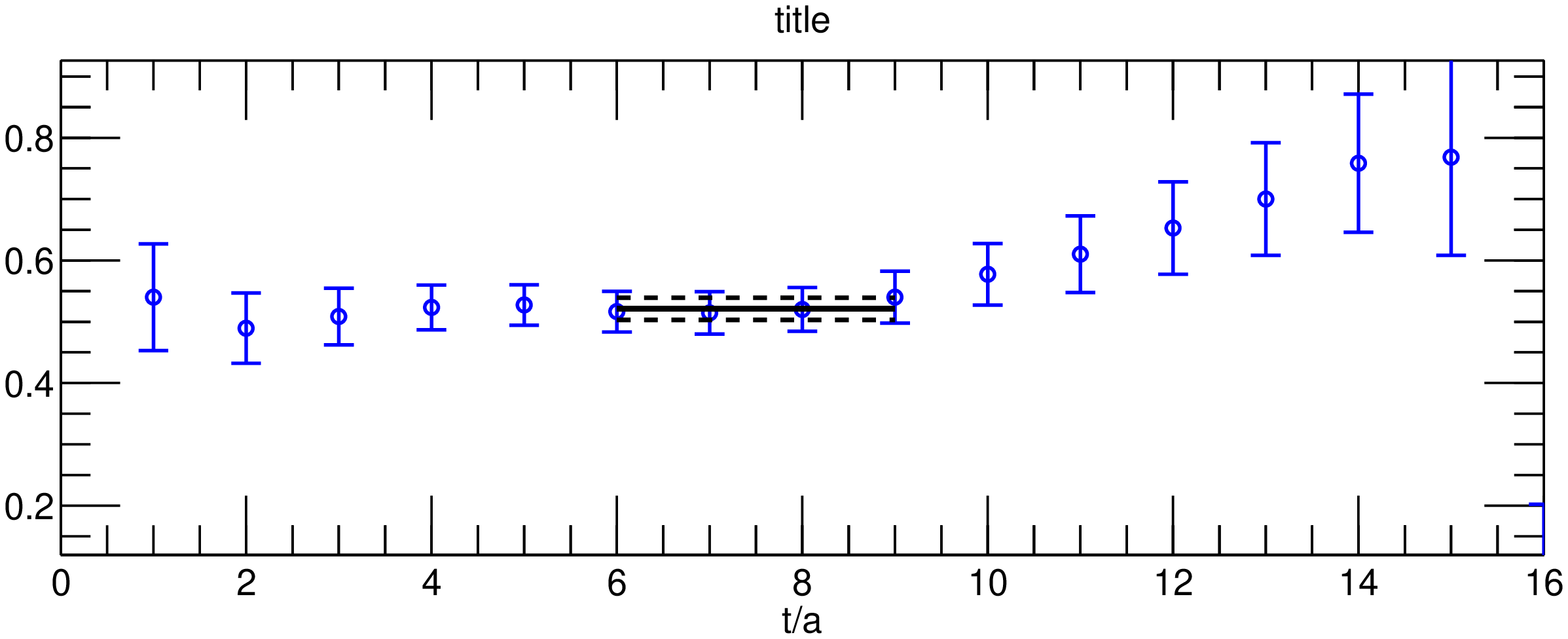}
\hspace{.5cm}
\epsfig{scale=.19,angle=-90,file=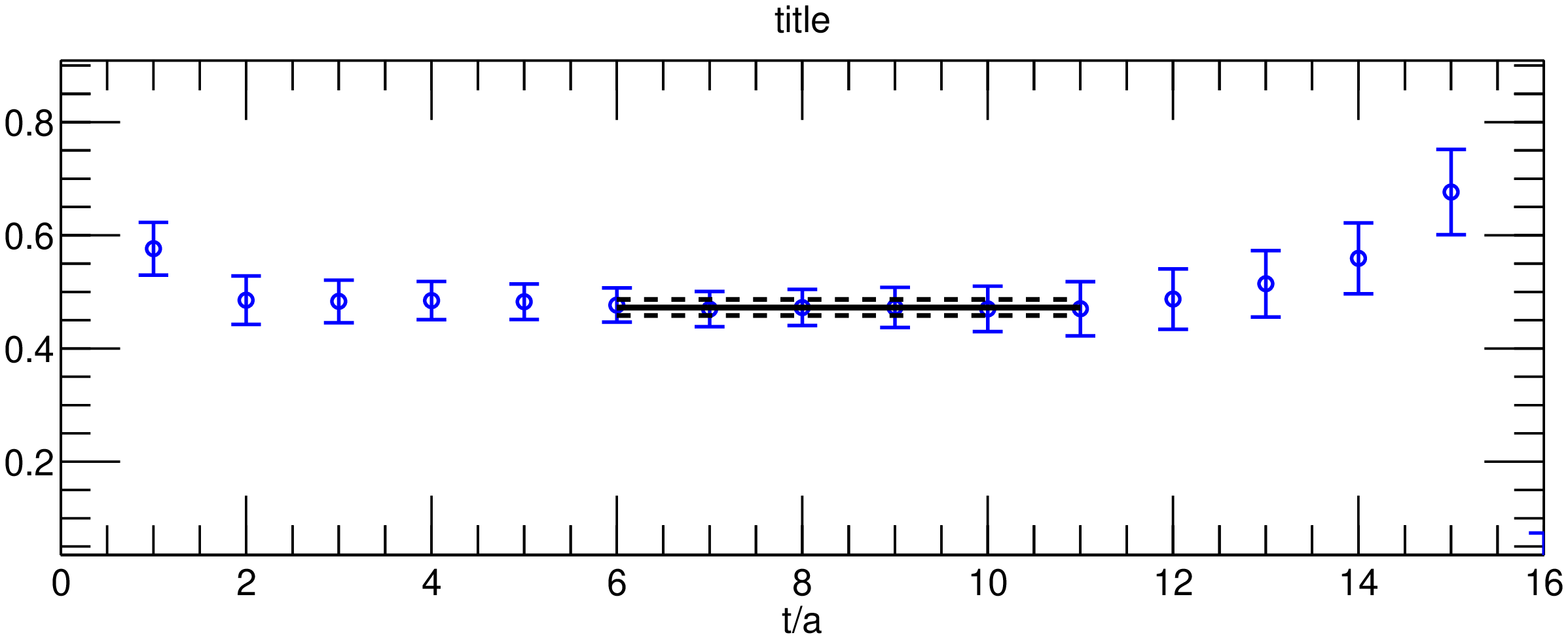}
\caption{Example of a typical result for $R_3$ (cf.
eq.~(\ref{eq:ratios})) for the pion form factor for $am_l=0.01$
(left) and $am_l=0.02$ (right). Here the source momentum is 
$|ap|=0.39$ and the sink momentum is $0.1$.
} \label{fig:comp_rat_pion}}
%%%%%%%%%%%%%%%%%%%%%%%%%%%%%%%%%%%%%%%%%%%%%%%%%%%%%%%%%%%%%%%%%%%%%%%%%%%%%%%%
\subsection{The $K\to\pi$ Form Factor, \boldmath{$f^+_{K\pi}(0)$}}

We computed the $K_{\ell 3}$-form factor using  both the
conventional and our new approach for two values of the light
quark mass on 200 configurations with a separation of 10
trajectories. The conventional approach has been described in
section~\ref{subsec:conventional} above; for more details on the
analysis see refs.~\cite{Becirevic:2004ya,Dawson:2006qc,Antonio:2006ev}.
For this conventional calculation, we
average all correlation functions over results from two positions
of the propagator source, $(0,0,0,0)$ and $(8, 8, 8, 16)$. For the
new approach we  generate correlation functions using only a
single source, (0,0,0,0), but averaging over three equivalent
twisting angles, i.e. $\vec{\theta}_i=(\theta_i,0,0)$,
$(0,\theta_i,0)$ or $(0,0,\theta_i)$, with $i=\pi$ or $K$. The
numerical values of $\theta_\pi$ and $\theta_K$ obtained using
eq.\,(\ref{eq:twists}) are presented in table \ref{tab:kin}.

In order to choose the twisting angles which correspond to $q^2=0$
using equation (\ref{eq:twists}) we need to know what the masses
of the mesons are. We initially estimated these using a subset of
100 configurations from the same ensembles and using only one
position for the source of the propagator. The mean values
obtained in this way are $am_\pi^{0.01}=0.306$,
$am_K^{0.01}=0.399$, $am_\pi^{0.02}=0.374$ and
$am_K^{0.02}=0.430$, to be compared with those eventually
determined on the full ensemble and averaged over various
positions of the propagator source given in table\,\ref{tab:kin}.
The small differences in the central values, together with
discretization effects in the pion and kaon dispersion relation
lead to a small deviation in $q^2$ from 0. We summarize this
effect in table\,\ref{tab:kin} where the quoted values for $q^2$
are obtained using eq.\,(\ref{eq:mom_transfer}) with the
corresponding meson energies determined from fits to the
respective two point correlation function.

We present our results in the plots in fig.\,\ref{fig:result_kl3}.
The left-hand plot shows the data points which one obtains from
correlation functions with one meson at rest and the other with
momentum of magnitude $\,|\vec p\, |=0,{2\pi/ L}$ or
$\sqrt{2}\,(2\pi /L)$. Since in addition to $f_{K\pi}^0(q^2)$ we
also compute $f^0_{\pi K}(q^2)$, from equations
(\ref{eq:conv_mainformula})-(\ref{Eqn:F_define}) (modified in the
obvious way) we obtain results for the form factor at four
additional values of $q^2$ for each value of the quark mass.

\begin{table}
 \begin{center}
 \begin{tabular}{l|c|cc}
\hline\hline&&\\[-4mm]
&       &$am_l=0.02$    &$am_l=0.01$\\
\hline\hline&&\\[-3mm]
&  $am_\pi$ &0.3765(21) &0.3002(23)\\[1mm]
&  $am_K$   &0.4312(20) &0.3942(20)\\[1mm]
\hline&&&\\[-4mm]
\multirow{5}{*}{$f_{K\pi}$} & ${\theta}_\pi$    &0.838
        &1.315\\[1mm]
&  ${\theta}_K$ &0.963      &1.714\\
    &$q^2_{\rm max}$&        0.00299(5)    &0.00883(20)\\
    &$q^2_{\theta_\pi}$&   -0.00012(4)   &0.00016(17)\\
    &$q^2_{\theta_K}$&     -0.00017(5)   &0.00017(23)\\[2mm]
\hline&&&\\[-4mm]
\multirow{1}{*}{$f_{\pi\pi}$} & ${\theta}_\pi$
    &0,\,1.6,\,2.3&0,\,1.6,\,2.3\\[2mm]
\hline\hline

 \end{tabular}\caption{Summary of kinematical parameters for
the studies involving twisted boundary conditions.}\label{tab:kin}
 \end{center}
\end{table}

\FIGURE{
\begin{minipage}{.48\linewidth}
\psfrag{xlabel}[t][t][1][0]{$(aq)^2$}
\psfrag{ylabel}[b][b][1][0]{$f_{K\pi}^0(aq)^2$}
\epsfig{scale=.28,angle=-90,file=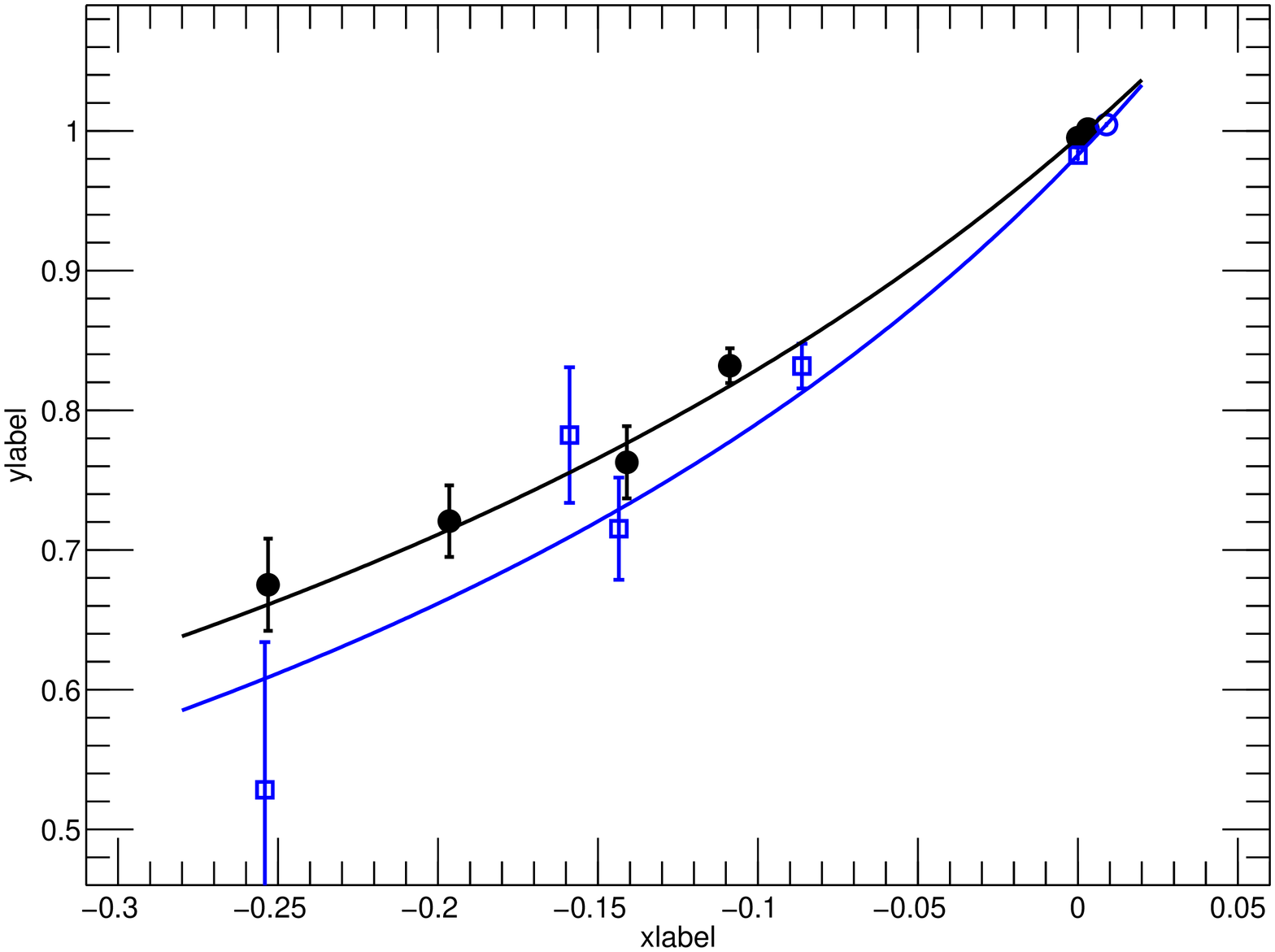}
\end{minipage}
\begin{minipage}{.48\linewidth}
\psfrag{xlabel}[t][t][1][0]{$(aq)^2$}
\psfrag{ylabel}[b][b][1][0]{$f_{K\pi}^0(aq)^2$}
\epsfig{scale=.28,angle=-90,file=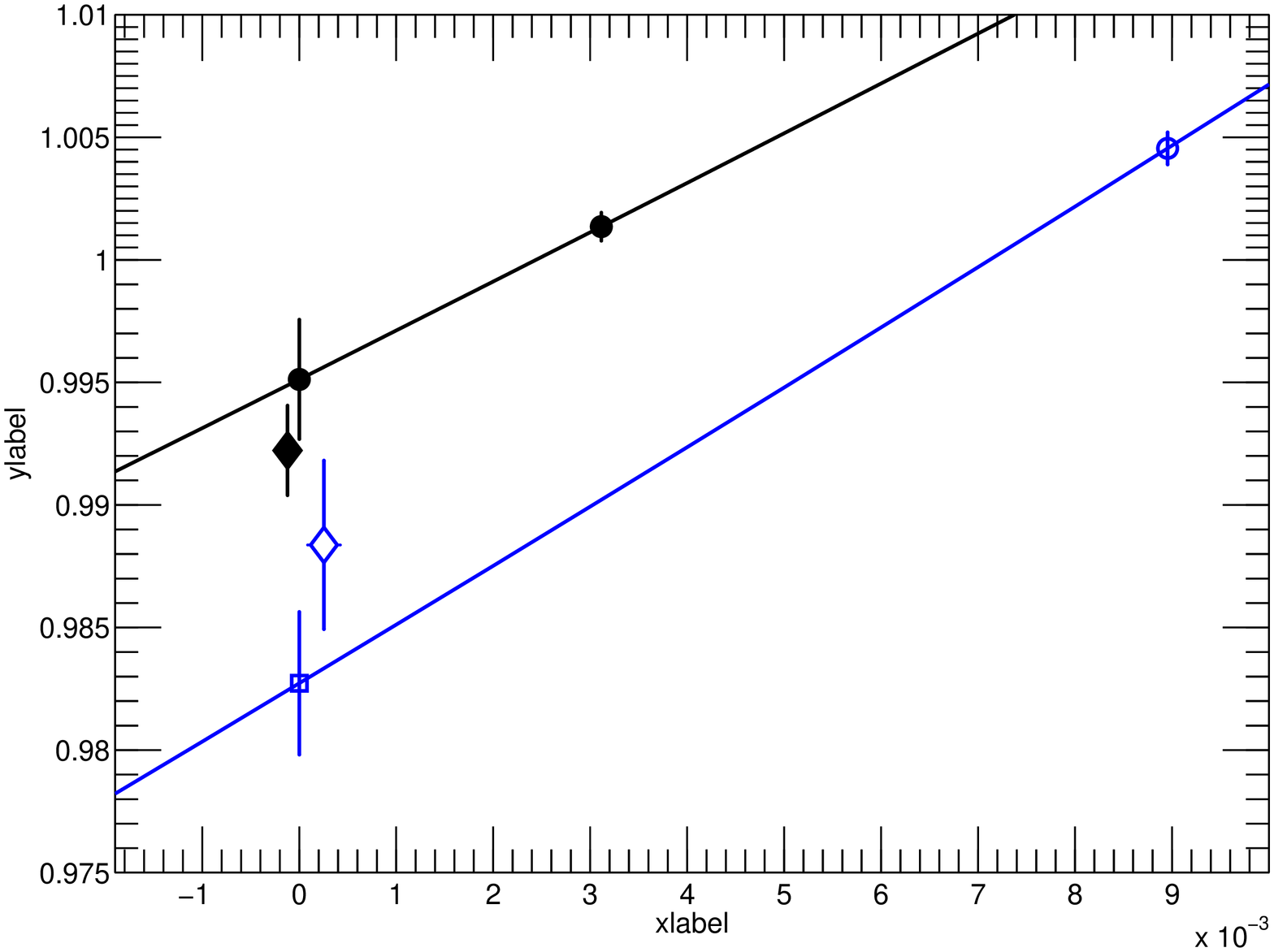}
\end{minipage}
\caption{Results for the form factor ($am_l=0.02$ as full black circles
 and $am_l=0.01$ as blue squares).
Left: All data points entering the conventional approach. Right:
Zoom which shows the data points for both
the new (diamonds) and the conventional approach at $q^2=0$ and the data
points at $q^2_{\rm max}$.}\label{fig:result_kl3}
} The corresponding values of $q^2$ are determined by using the
pion and kaon masses obtained from fits to two-point functions as
input to the continuum dispersion relation. The curves represent a
fit using the pole-dominance ansatz
\begin{equation}\label{eq:Kl3_PD_ansatz}
f^0_{K\pi}(q^2)=\frac{f_{K\pi}^0(0)}{1-q^2/M^2}\,,
\end{equation}
where $M$ is a parameter fitted from the data.

The right-hand plot shows a zoom into the region around $q^2=0$.
The two data points at $q^2>0$ correspond to the results for
$q^2_{\rm max}$ for which the pion and kaon are both at rest; they
can be identified by their strikingly small errors. We also
display the results for $f_{K\pi}^0(q^2=0)$ obtained from the pole
fit (\ref{eq:Kl3_PD_ansatz}) at each quark mass. In addition the
right plot of fig.~\ref{fig:result_kl3}
 contains the results from the
new approach.

We obtain the following results using the conventional and the new
approach:
\begin{center}
\begin{tabular}{cccc}
        &$am_l$&    0.02    &0.01\\
\hline\hline&&&\\[-4mm]
\multirow{2}{*}{$f_{K\pi}^0(0)$}
    &conventional &    0.9951(24)&0.9827(29)\\
        &new          &    0.9926(34)&0.9884(34)\\
\end{tabular}
\end{center}
The results for $f_{K\pi}(0)$ as determined from the conventional
and the new approaches do not agree exactly but the discrepancy is
statistically not significant. The size of the statistical errors
is similar in the two approaches, which is an important condition
for establishing the new technique. The main motivation for the
direct approach advocated in this paper is to avoid the need for
an ansatz with which to perform the $q^2$ interpolation and this
is apparently achieved without significantly inflating the error.
In addition, it should be stressed that the points at negative
$q^2$ are obtained with $|\,\vec p\,|=2\pi/L$ and $\,|\vec
p\,|=\sqrt{2}\,2\pi/L$, so that with $L/a=16$ one may have
concerns about the size of the lattice artefacts at these momenta.
Our results show that both concerns can now be eliminated by using
the new approach.

As stated in the introduction, this is an exploratory study in
which we investigate the feasibility of the method rather than aim
for the ultimate physical results. In particular it will be
important to check the precision of the direct approach as the
mass of the light quark is reduced and/or the volume is increased.
Reducing the light quark mass leads to an increase of $q^2_{\rm
max}$ and therefore the value for the form factor $f_{K\pi}(0)$
will be more susceptible to the choice for the interpolation in
$q^2$ in the conventional approach. Simulating in larger volumes,
in addition to reducing the finite-volume corrections, enables
smaller Fourier momenta (with components which are integer
multiples of $2\pi/L$) to be reached, so that the data points at
non-vanishing momentum in the conventional approach move closer to
$q^2=0$ and thus better constrain the interpolation.

%%%%%%%%%%%%%%%%%%%%%%%%%%%%%%%%%%%%%%%%%%%%%%%%%%%%%%%%%%%%%%%%%%%%%%%%%%%%%%%%
\subsection{The Electromagnetic Form Factor of the Pion, \boldmath{$f_{\pi\pi}(q^2)$}}

We compute the pion's electromagnetic form factor for two
values of the light quark mass $am_l=0.02$ and $am_l=0.01$ on 200
configurations separated by 10 trajectories in Monte Carlo time.
We generate pion three-point functions $C_{\pi\pi}$ with all
possible mutual combinations of the twisting angles $\theta_{1,2}$
 as given in table \ref{tab:kin}
(cf. also fig.\,\ref{fig:threepoint}), with the twist
being applied only in one direction (i.e.
$\vec\theta_{1,2}=(\theta_{1,2},0,0)$). After projecting on the
Fourier momenta of magnitudes $|\vec p_{\rm FT}|=0,$ $2\pi/L,$ and
$\sqrt{2}\,(2\pi/L)$ for the initial state, we are thus able to
generate data points for the form factor in the entire range from
$q^2=0$ to approximately $1\,{\rm GeV}^2$. We average the results
obtained for degenerate values of $q^2$ (see
eq.\,(\ref{eq:mom_transfer})\,). A simplification compared to the
$K_{\ell 3}$ scalar form factor is that we do not need $Z_V$;
current conservation implies that $f_{\pi\pi}(0)=1$ and this
provides the required normalization.

As was noted in ref.\,\cite{Brommel:2006ww}, for large values of the
initial and/or final pion's momentum the argument of the square
root in $R_3$ may become negative due to statistical
fluctuations\footnote{Since the induced pion and kaon momenta in
the determination of the $K_{\ell 3}$ form factor with twisted
boundary conditions are small this problem does not occur there.}.
To avoid this problem we follow ref.\,\cite{Brommel:2006ww} and
consider the two-point functions at smaller times, making the
replacement,
\begin{equation}
C_\pi(T/2,\vec p)\to\frac{C_\pi(T/2-t_{\rm shift},\vec p)}
        {\cosh(E_\pi(\vec p)t_{\rm shift})},
\end{equation}
and we find that $t_{\rm shift}/a = 7$ gives the best
results. This removes the occurrence of negative arguments in the
square root in (\ref{eq:ratios}) for most of the values of $\vec
p$ used in this paper. The remaining kinematic points for which we
still find negative arguments for the square root in the range of
time-slices we wish to fit to are removed from our analysis.

The results are shown in  fig.\,\ref{fig:twisted_pff}.
The vertical lines correspond to those values of
$q^2=2m_\pi(m_\pi-\sqrt{m_\pi^2+(2\pi/L)^2})$ below which form
factors cannot be computed with the conventional periodic boundary
conditions. \FIGURE{ \psfrag{8}[c][c][1][0]{$O$}
\hspace*{3mm}\begin{minipage}{.45\linewidth}
\psfrag{xlabel}[t][t][1][0]{\small$-q^2/{\rm GeV}^2$}
\psfrag{ylabel}[c][t][1][0]{\small$f_{\pi\pi}(q^2)$}
\epsfig{scale=.28,angle=-90,file=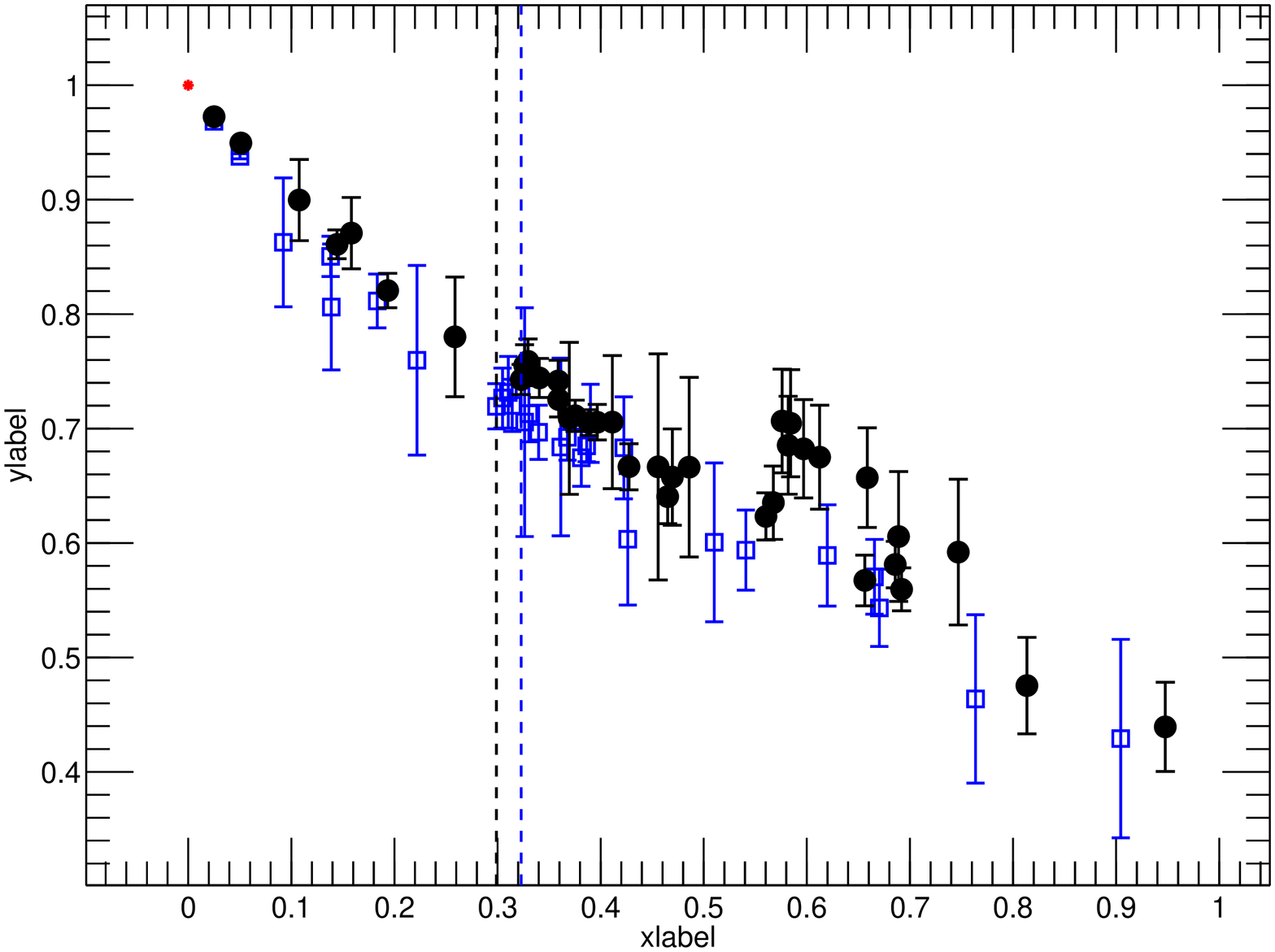}
\end{minipage}
\hspace*{.2cm}
\begin{minipage}{.45\linewidth}
\psfrag{xlabel}[t][t][1][0]{\small$-q^2/{\rm GeV}^2$}
\psfrag{ylabel}[c][t][1][0]{\small$f_{\pi\pi}(q^2)$}
\epsfig{scale=.28,angle=-90,file=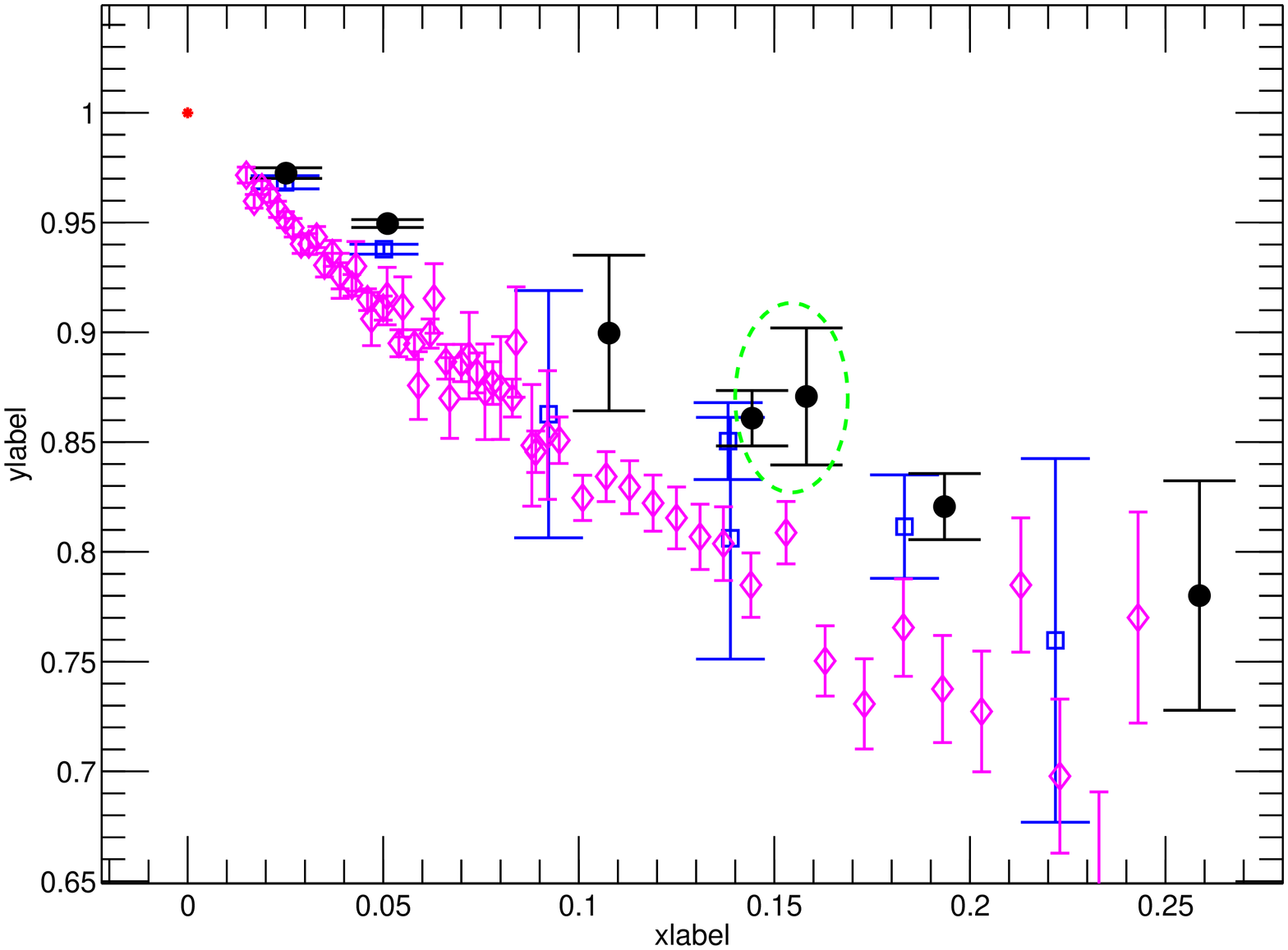}
\end{minipage}
\caption{Left: All the numerical results for the pion form factor
    ($m_l=0.02$ circles and $m_l=0.01$ squares).
    Right: Expanded view of low $-q^2$-region, now also including
    experimental results (diamonds)
    \cite{Amendolia:1986wj,Dally:1982zk} and $f_{\pi\pi}(0)= 1$ (star).}
    \label{fig:twisted_pff}
} The first observation is that our method works very well and
that we can indeed compute the pion form factor in the low
$|q|^2$-regime.

We find that the size of the errors correlates very well with the
magnitudes of the momenta of the initial and final state pions.
This agrees with the observations of ref.\,\cite{Flynn:2005in}
where the statistical noise as a function of the induced meson
momentum was investigated numerically in two-point functions and
found to increase with the momentum. Similar values of $q^2$ may
be obtained by pions with very different momenta, and we find that
the errors in the form factors are generally smaller if the pions'
momenta are smaller. As an example consider the two results for
the pion form factor at the two neighbouring values of
$-q^2\approx0.14\,{\rm GeV}^2$ and $0.16\,{\rm GeV}^2$ for
$am_l=0.02$ which we indicate by the dashed circle in the right
hand plot in fig.\,\ref{fig:twisted_pff}.
In  both cases the pion at the sink has a small momentum,
$|ap|=0$ and $0.1$, respectively. At the source however the pion's momentum
is $0.25$ in the one case (smaller error bar) and $0.39$ in the other case
(larger error bar).
In addition, in some cases we achieve a
reduction in the error by averaging results for the form factor at
degenerate values of the momentum transfer.
\subsection{Fits to the Pion Form Factor}\label{sec:Chi}
Fig.\,\ref{fig:twisted_pff} contains the main results of our study
for the pion form factor. They are very encouraging, clearly
demonstrating the feasibility of the method. Ultimately of course,
after performing our large scale simulation, we will wish to
compare our results with experimental measurements, but at this
stage we can only perform some rudimentary analyses. In particular
we investigate various fit-ans\"atze for our present data with the
aim of extracting the pion's charge radius,
\begin{equation}
\langle r_\pi^2 \rangle =6 {d f_{\pi\pi}\over d q^2}\Big|_{q^2=0}\,,
\end{equation}
which has also been measured in various experiments.
For a qualitative comparison of our data to experiment at low values of
$|q^2|$ we have
added the experimental data of refs.\,\cite{Amendolia:1986wj} and
\cite{Dally:1982zk} to the r.h.s. plot in
fig.\,\ref{fig:twisted_pff}. We note, that there also exist new measurements
of the Pion form factor at larger values of $|q^2|$ \cite{Tadevosyan:2006yd,
Horn:2006tm}.

One approach is to use a pole-dominance (PD) ansatz of the form:
\begin{equation}\label{eq:PD_ansatz}
f_{\pi\pi,{\rm PD}}(q^2)=\frac{n}{1-{q^2}/{M^2_{\rm PD}(m_\pi)}}\, ,
\end{equation}
where the pole mass is related to the pion charge radius by
$M_{\rm PD}^2=6/\langle r_\pi^2\rangle_{\rm PD}$. Note that for
lattice data $n=1$, whereas for experimental data one either also
sets $n=1$ or leaves it as a free parameter due to uncertainties
in the overall normalization (see \cite{Bijnens:2002hp} for
example). In order to extract the physical value of the charge
radius from lattice data one determines $M_{\rm PD}^2(m_\pi)$ for
various values of $m_\pi$ and extrapolates the pole mass to the
chiral limit (cf. e.g. \cite{Brommel:2006ww}). Since we only have
data for two values of the pion mass we will not extrapolate
$M_{\rm PD}^2(m_\pi)$ here and merely compare the fit results at
fixed pion mass with the ones from the polynomial ans\"atze,
\begin{equation}\label{eq:poly_ansatz}
f_{\pi\pi,{\rm lin}}(q^2)=1+\frac 16 \langle r_\pi^2\rangle_{\rm lin}\, q^2\;\;
{\rm and}\;\;
f_{\pi\pi,{\rm quad}}(q^2)=1+\frac 16 \langle r_\pi^2\rangle_{\rm quad}\, q^2
    + c_\pi \,q^4\,.
\end{equation}

Other ans\"atze are guided by the prediction of the chiral
effective theory
\cite{Weinberg:1978kz,Pagels:1974se,Gasser:1983yg,Gasser:1984gg}
where the pion form factor is a well-studied observable
\cite{Gasser:1984ux,Bijnens:1987dc,Bijnens:2002hp}. Here we quote
the result at next-to-leading order (NLO)
\cite{Gasser:1984ux,Bijnens:1987dc},
\begin{equation}\label{eq:fpipi_NLO}
f_{\pi\pi,{NLO}}(q^2)=1+\frac{1}{f_\pi^2}\left\{
    2L_9^rq^2+2\tilde{\mathcal{H}}(m_\pi^2,q^2)+\tilde{\mathcal{H}}(m_K^2,q^2)
    \right\}
\end{equation}
where $f_\pi=92.4(5)(3)$MeV is the physical pion decay constant
\cite{Yao:2006px}, $L_9^r$ the only low energy constant relevant
at this order of the effective theory and we define\footnote{ Note
that the functions $\tilde{\mathcal{H}}$ and $\tilde \nu$ are
 a slight modification of the functions
$\mathcal{H}$  and $\nu$ in \cite{Bijnens:2002hp}.}
\begin{eqnarray}
\tilde{\mathcal{H}}(m^2,q^2)&=&
q^2\left(\frac{5}{576\pi^2}-\frac{1}{192\pi^2}\log(m^2/\mu^2)\right)
    -\frac{1}{24\pi^2}m^2\nonumber\\[2mm]
    &&+\left(\frac{1}{96\pi^2}\frac{m^2}{q^2}-\frac{1}{384\pi^2}\right)
    \tilde\nu(m^2,q^2)\log\left[
    \frac{2m^2-q^2-\tilde\nu(m^2,q^2)}{2m^2-q^2+\tilde\nu(m^2,q^2)}\right]\,,
\end{eqnarray}
with $\tilde\nu(m^2,q^2)=\sqrt{q^4-4q^2m^2}$.

We have carried out the following fits:
\begin{itemize}
\item[A)] $\langle r_\pi^2 \rangle$ from fits of (\ref{eq:PD_ansatz}),
    (\ref{eq:poly_ansatz}) and (\ref{eq:fpipi_NLO}) to the lattice data
    at the unphysical values of the quark mass $am_l=0.02$ and
    $0.01$;
\item[B)] $\langle r_\pi^2 \rangle_\chi$ and $L_9^r$, where the subscript
    $\chi$ indicates that the chiral limit has been taken,
    from  global fits to the results at the two pion masses using
    the NLO expression (\ref{eq:fpipi_NLO}),
\item[C)] $\langle r_\pi^2 \rangle_\chi$ from global fits to the lattice results
    at the two pion masses using the phenomenologically motivated ansatz
    \begin{equation}\label{eq:NLO_pheno}
    f_{\pi\pi,{\rm pheno}}(q^2)=f_{\pi\pi,NLO}(q^2)+c_1q^2m_\pi^2
        +c_2 q^2m_K^2+c_3q^4\,.
    \end{equation}
\end{itemize}

The results are summarized in fig.\,\ref{fig:finite_mq_res} and
table \ref{tab:chi_res}. The plots show how the result of each fit
changes under a variation of the range in $q^2$ for the various
fit functions discussed above. We keep the lowest point in the fit
to the $q^2$ behaviour fixed at $q^2=0$ and vary the upper
limit. When we quote the lattice results extrapolated to the
chiral limit we mean the point defined by $m_\pi=140$ MeV and
$m_K=494$ MeV. For the final results which we quote in table
\ref{tab:chi_res} we chose the upper limit of the fit range to be
$\approx 0.1\,{\rm GeV}^2$ in the case of the fits A) and B)  and
about $0.35\,{\rm GeV}^2$ for fit C). \FIGURE{
\begin{minipage}{.45\linewidth}
\psfrag{ylabel1}[b][b][1][0]{$\langle r_\pi^2 \rangle_{\rm lin}$}
\psfrag{ylabel2}[b][b][1][0]{$\langle r_\pi^2 \rangle_{\rm quad}$}
\psfrag{ylabel3}[b][b][1][0]{$\langle r_\pi^2 \rangle_{\rm PD}$}
\psfrag{xlabel}[t][b][1][0]{$-q^2$}
\epsfig{scale=.27,angle=-90,file=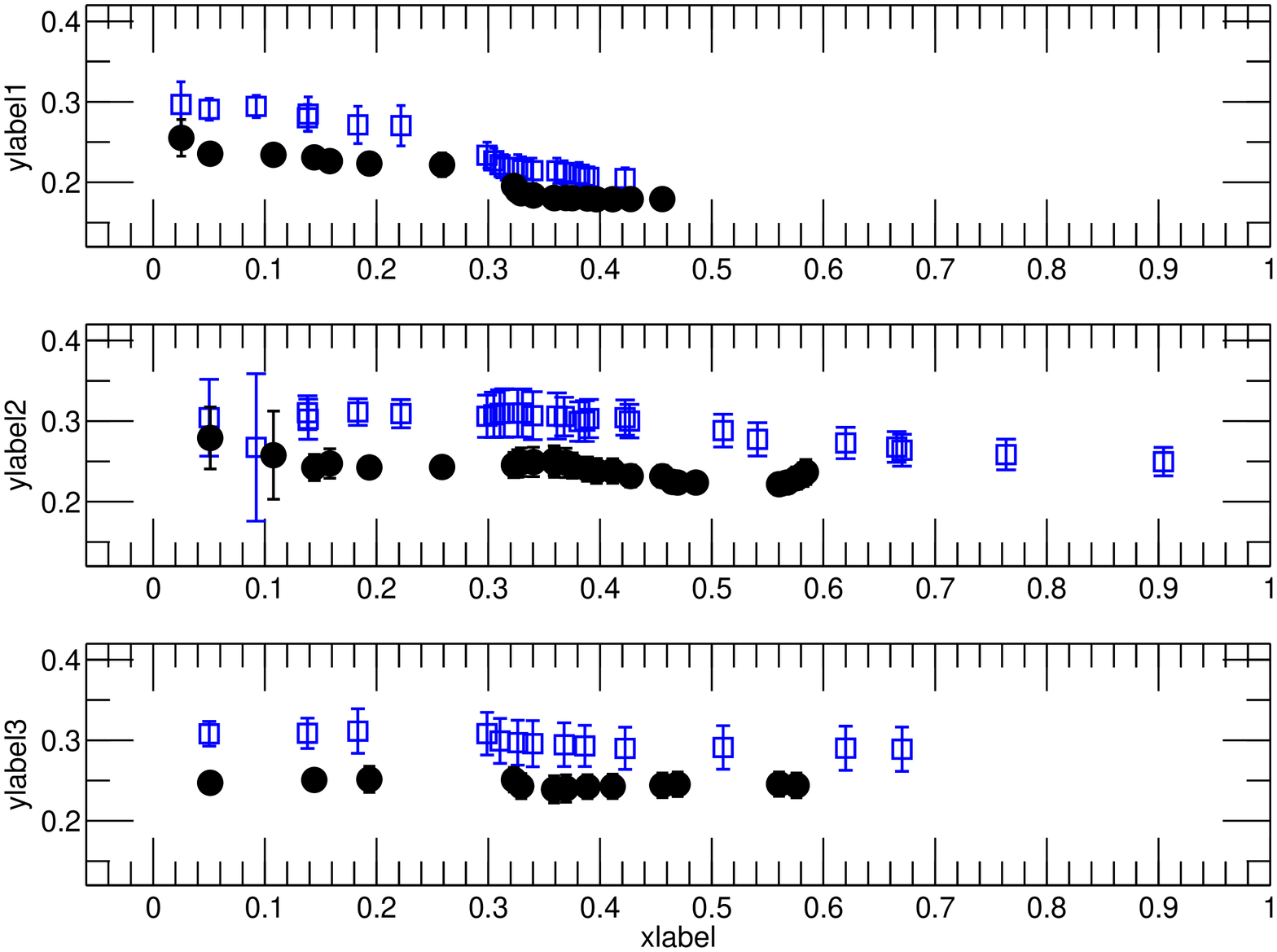}
\end{minipage}
\hspace*{2mm}
\begin{minipage}{.45\linewidth}
\psfrag{ylabel1}[b][b][1][0]{$\langle r_\pi^2\rangle_{\rm NLO}$}
\psfrag{ylabel2}[b][b][1][0]{$\langle r_\pi^2\rangle_{\rm NNLO}$}
\psfrag{xlabel}[t][b][1][0]{$-q^2$}
\epsfig{scale=.27,angle=-90,file=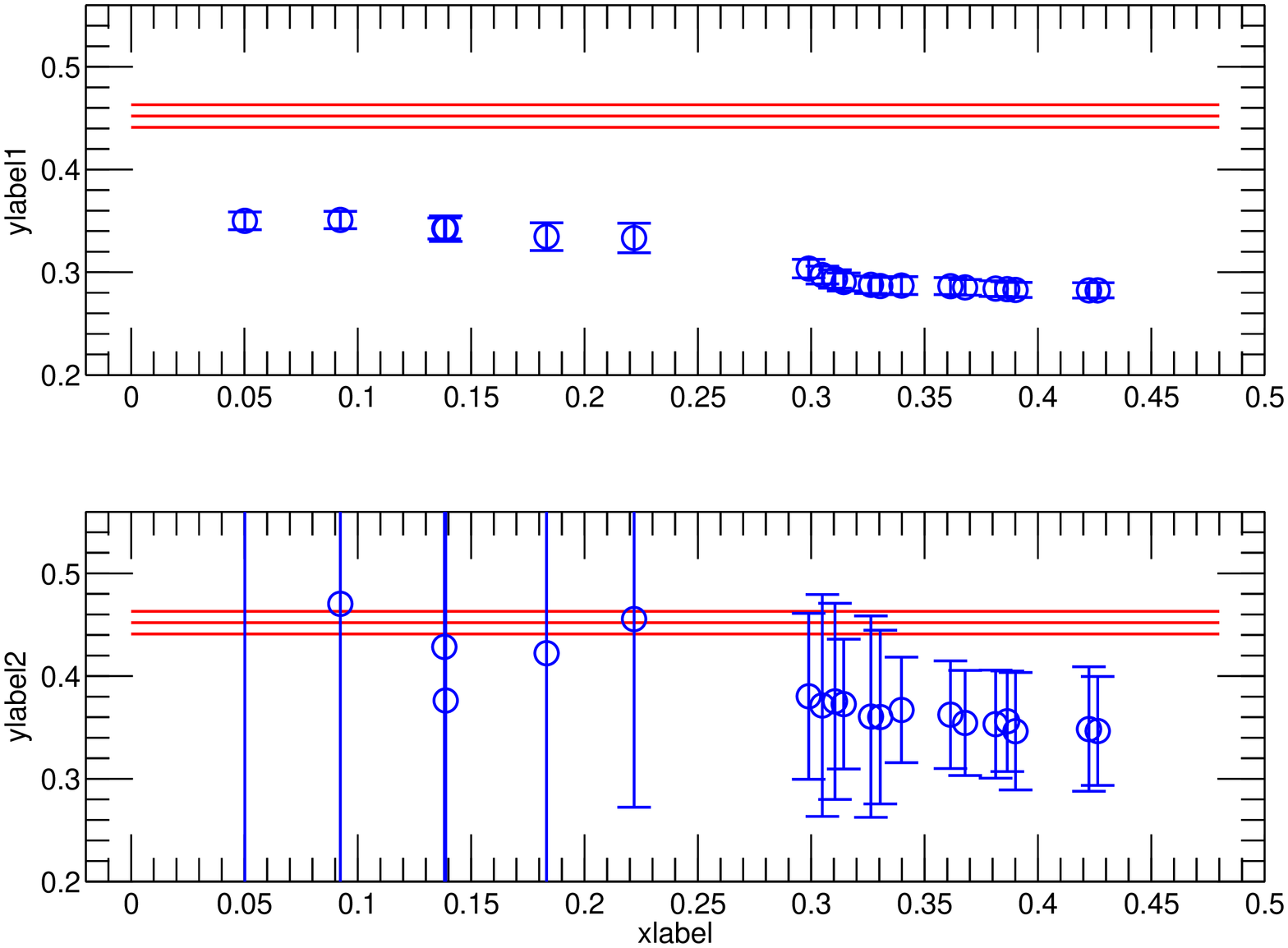}
\end{minipage}
\caption{The plots show results from fits to the pion form factor
with various
    fit functions. Left: Results for the charge radius at
    $am_l=0.02$ (black circles) and $am_l=0.01$ (blue squares) for
    linear, quadratic and pole dominance ansatz, respectively.
    Right: Fits B) and C) compared to the PDG value for the
    charge radius.
    }\label{fig:finite_mq_res}
}
\begin{table}
\begin{center}
 \begin{tabular}{cccccccc}
 \hline\hline\\[-3mm]
&\multicolumn{2}{c}{fit}   &$\langle r_\pi^2\rangle|_{0.01}/{\rm fm}^2$&
    $\langle r_\pi^2\rangle|_{0.02}/{\rm fm}^2$\\[1mm]
 \hline&&\\[-4mm]
 \multirow{3}{*}{A)}&lin. &eq.~(\ref{eq:poly_ansatz})&0.294(13) &0.234(10)\\[2mm]
            &quad.&eq.~(\ref{eq:poly_ansatz})&0.311(16) &0.242(13)\\[2mm]
            &PD &eq.~(\ref{eq:PD_ansatz})&0.297(28) &0.239(17)\\[2mm]
 \hline\hline\\
 \end{tabular}\\[8mm]
 \begin{tabular}{ccccc}
 \hline\hline\\[-3mm]
 fit  &$\langle r_\pi^2\rangle_\chi/{\rm fm}^2$&$L_9^r(m_\rho)|_{\chi}$\\[1mm]
 \hline&&\\[-4mm]
 B)&    0.351(8)  &0.0050(2)  \\[2mm]
 C)&    0.37(5)   &0.006(2)\\[2mm]
 \hline\hline\\
 \end{tabular}
\end{center}
\caption{Results of pion form factor fits.}\label{tab:chi_res}
\end{table}

All the fits in case A) for the same quark mass are compatible
within errors for small values of $|q^2|$. The linear fit ansatz
starts deviating from the results of the quadratic and PD ansatz
between $0.2$ and $0.3\,{\rm GeV}^2$. The second order polynomial
ansatz and the PD ansatz turn out to be stable and rather constant
in the error over a large range of $q^2$. We observe a smaller
error for the result of the linear fit for small values of
the momentum transfer and in this region this is therefore the preferred
ansatz.

For case B) we used eq.~(\ref{eq:fpipi_NLO}) in a global fit over
the results for $am_l=0.01$ and $am_l=0.02$ and try to obtain a
prediction for the charge radius in the chiral limit or
equivalently for the low energy constant $L_9^r$ (we set
$\mu=m_\rho=0.77$\,GeV). For the results which we quote we choose
values extracted from the fit including all data points in the
interval $q^2\in [0,0.1]{\rm GeV^2}$.  The results are plotted in
the upper r.h.s. plot of fig.\,\ref{fig:finite_mq_res}. The blue
circles correspond to our lattice results for the charge radius
and we indicate the PDG value $\langle r_\pi^2\rangle=0.452(11){\rm fm}^2$
\cite{Yao:2006px} by the horizontal line with error band. Our
results disagree very significantly with the PDG value indicating
that our quark masses are too heavy to be described by NLO chiral
perturbation theory. This conclusion is reinforced by the fact
that the values for $L_9^r$ as determined independently at
$am_l=0.01$ and $am_l=0.02$ are significantly different:
$L_9^{r}|_{am_l=0.01}=0.0056(3)$ and
$L_9^{r}|_{am_l=0.02}=0.0048(2)$. For comparison we also quote the
value  $L_9^{r}=0.00593(43)$ determined by Bijnens \cite{Bijnens:2002hp}
using experimental data together with 
next-to-next-to-leading order chiral perturabation theory.

The results for fit C) are summarized in the lower r.h.s. plot in
fig.\,\ref{fig:finite_mq_res}. This global fit of the
phenomenologically motivated ansatz (\ref{eq:NLO_pheno}) leads to
results  for the charge radius with larger errors.
 The results which we quote in table \ref{tab:chi_res}
correspond to the values extracted from the fit including all data
points in the interval $q^2\in [0,0.35]{\rm GeV^2}$.

%%%%%%%%%%%%%%%%%%%%%%%%%%%%%%%%%%%%%%%%%%%%%%%%%%%%%%%%%%%%%%%%%%%%%%%%%%%%%%%%
\section{Conclusions}\label{sec:concs}
%%%%%%%%%%%%%%%%%%%%%%%%%%%%%%%%%%%%%%%%%%%%%%%%%%%%%%%%%%%%%%%%%%%%%%%%%%%%%%%%
In this paper we have demonstrated the feasibility of using
partially twisted boundary conditions to compute weak and
electromagnetic form factors. The technique allows the form
factors to be evaluated for any choice of momentum transfer, $q$,
within the range of hadronic momenta such that lattice artefacts
are small. We have illustrated the techniques for two
phenomenologically interesting quantities, the evaluation of the
form factors for $K_{\ell 3}$ decays directly at $q^2=0$ (thus
avoiding the need for an interpolation in $q^2$) and of the
electromagnetic form factor of the pion at small momentum
transfers (thus enabling us to evaluate the pion's charge radius
without extrapolation of the results from large values of $-q^2$).

The computations described here were intended to demonstrate the
proof of concept, and as such were limited both in statistics and
in the lattice parameters. In particular we only compute the form
factors for two values of $m_{u,d}$ and hence any investigation of
the chiral behaviour is restricted. The next step will be to
implement the technique in a large scale simulation which will
enable us to overcome the limitations of this feasibility study
and to obtain results with the systematic uncertainties under
control.

{\bf Acknowledgements:} We warmly thank Johan Bijnens and Giovanni
Villadoro for very helpful discussions.

We gratefully acknowledge the QCDOC design team for developing the
QCDOC supercomputer used in these calculations. The development
and the computers used in this calculation were funded by the U.S.
DOE grant DE-FG02-92ER40699, PPARC JIF grant PPA/J/S/1998/00756
and by RIKEN. We thank the University of Edinburgh, PPARC, RIKEN,
BNL and the U.S. DOE for providing these facilities. This work was
supported by PPARC grants PPA/G/O/2002/00465,
PPA/G/S/\-2002/00467, PP/D000211/1, \linebreak PP/D000238/1.

%%%%%%%%%%%%%%%%%%%%%%%%%%%%%%%%%%%%%%%%%%%%%%%%%%%%%%%%%%%%%%%%%%%%%%%%%%%%%%%%
\bibliographystyle{JHEP}
\bibliography{tw_ff}

\end{document}